\documentclass[12pt,preprint]{aastex} %\documentclass{emulateapj}

\renewcommand{\cite}{\citealp} %%%\makeatletter

\usepackage{longtable}

\shortauthors{M. Dall'Ora et al.}

\begin{document}

%% LaTeX will automatically break titles if they run longer than %% one
%line.However, you may use \\ to force a line break if %% you desire.

\title{The Type IIP Supernova 2012aw in M95: hydrodynamical modelling of the photospheric phase from accurate spectrophotometric monitoring.}

%% Use \author, \affil, and the \and command to format %% author and affiliation
%information. %% Note that \email has replaced the old \authoremail command %%
%from AASTeX v4.0. You can use \email to mark an email address %% anywhere in the %paper, not just in the front matter. %% As in the title, use \\ to force line %breaks.

\author{M. Dall'Ora \altaffilmark{1}} 
\author{M.~T. Botticella\altaffilmark{1}}
\author{M.~L. Pumo\altaffilmark{2}} 
\author{L. Zampieri\altaffilmark{2}}
\author{L. Tomasella\altaffilmark{2}} 
\author{G. Pignata\altaffilmark{3}}
\author{A.~J. Bayless\altaffilmark{4}} 
\author{T.~A. Pritchard\altaffilmark{5}}
\author{S. Taubenberger\altaffilmark{6}} 
\author{R. Kotak\altaffilmark{7}} 
\author{C. Inserra\altaffilmark{7}} 
\author{M. Della Valle\altaffilmark{1}} 
\author{E. Cappellaro\altaffilmark{2}} 
\author{S. Benetti\altaffilmark{2}} 
\author{S. Benitez\altaffilmark{6}} 
\author{F. Bufano\altaffilmark{3}} 
\author{N. Elias-Rosa\altaffilmark{8}} 
\author{M. Fraser\altaffilmark{7}} 
\author{J.~B. Haislip\altaffilmark{9}} 
\author{A. Harutyunyan\altaffilmark{10}} 
\author{D.~A. Howell\altaffilmark{11,12}} 
\author{E.~Y. Hsiao\altaffilmark{13}} 
\author{T. Iijima\altaffilmark{2}} 
\author{E. Kankare\altaffilmark{14}} 
\author{P. Kuin\altaffilmark{15}} 
\author{J.~R. Maund\altaffilmark{7,19}} 
\author{A. Morales-Garoffolo\altaffilmark{8}}
\author{N. Morrell\altaffilmark{13}}
\author{U. Munari\altaffilmark{2}} 
\author{P. Ochner\altaffilmark{2}} 
\author{A. Pastorello\altaffilmark{2}} 
\author{F. Patat\altaffilmark{16}} 
\author{M.~M. Phillips\altaffilmark{13}} 
\author{D. Reichart\altaffilmark{9}} 
\author{P.~W.~A. Roming\altaffilmark{4,5}} 
\author{A. Siviero\altaffilmark{17}} 
\author{S.~J. Smartt\altaffilmark{7}} 
\author{J. Sollerman\altaffilmark{18}} 
\author{F. Taddia\altaffilmark{18}} 
\author{S. Valenti\altaffilmark{11,12}} 
\author{D. Wright\altaffilmark{7}}

\altaffiltext{1}{INAF, Osservatorio Astronomico di Capodimonte, Napoli, Italy}  \altaffiltext{2}{INAF, Osservatorio Astronomico di Padova, 35122 Padova, Italy} \altaffiltext{3}{Departamento de Ciencias Fisicas, Universidad Andres Bello, Avda. Republica 252, Santiago, Chile}
\altaffiltext{4}{Southwest Research Institute, Department of Space Science, 6220 Culebra Road, San Antonio, TX 78238, USA}
\altaffiltext{5}{Department of Astronomy \& Astrophysics, Penn State University, 525 Davey Lab, University Park, PA 16802, USA}
\altaffiltext{6}{Max-Planck-Institut f\"ur Astrophysik, Karl-Schwarzschild-Str. 1, D-85741 Garching, Germany}
\altaffiltext{7}{Astrophysics Research Centre, School of Mathematics and Physics, Queen's University Belfast, Belfast, BT7 1NN, UK}
\altaffiltext{8}{Institut de Ci\`encies de l'Espai (IEEC-CSIC), Facultat de Ci\`encies, Campus UAB, 08193 Nellaterra, Spain}
\altaffiltext{9}{Department of Physics and Astronomy, University of North Carolina at Chapel Hill, 120 E. Cameron Ave., Chapel Hill, NC 27599, USA}
\altaffiltext{10}{ Fundaci\'{o}n Galileo Galilei - Telescopio Nazionale Galileo, Rambla Jos\'{e} Ana Fern\'{a}ndez P\'{e}rez 7, 38712 Bre\~{n}a Baja, TF - Spain}
\altaffiltext{11}{Las Cumbres Observatory Global Telescope Network, 6740 Cortona
Dr., Suite 102 Goleta, CA 93117, USA} 
\altaffiltext{12}{Department of Physics, University of California, Santa Barbara, Broida Hall, Mail Code 9530, Santa Barbara, CA 93106-9530, USA}
\altaffiltext{13}{Carnegie Observatories, Las Campanas Observatory, Colina El Pino, Casilla 601, Chile}
\altaffiltext{14}{Finnish Centre for Astronomy with ESO (FINCA), University of Turku, V\"ais\"al\"antie 20, FI-21500 Piikki\"o, Finland}
\altaffiltext{15}{Mullard Space Science Laboratory, Holmbury St. Mary, Dorking, Surrey RH5 6NT, UK}
\altaffiltext{16}{European Organization for Astronomical Research in the Southern Hemisphere (ESO), Karl-Schwarzschild-Str. 2, 85748, Garching b. München, Germany}
\altaffiltext{17}{Dipartimento di Fisica e Astronomia Galileo Galilei,
Universit\'a di Padova, vicolo dell'Osservatorio 3, 35122 Padova, Italy}
\altaffiltext{18}{The Oskar Klein Centre, Department of Astronomy, AlbaNova, SE-106 91 Stockholm, Sweden}
\altaffiltext{19}{Royal Society Research Fellow}

%%style, abstract will output a Received/Accepted line after the %% title andaffiliation information. No date will appear since the author %% does not havethis information. The dates will be filled in by the %% editorial office aftersubmission.

\begin{abstract} 

We present an extensive optical and near-infrared photometric and spectroscopic campaign of the type IIP supernova SN 2012aw. The dataset densely covers the evolution of SN 2012aw shortly after the explosion up to the end of the photospheric phase, with two additional photometric observations collected during the nebular phase, to fit the radioactive tail and estimate the $^{56}$Ni mass.  Also included in our analysis is the already published \textit{Swift} UV data, therefore providing a complete view of the ultraviolet-optical-infrared evolution of the photospheric phase. On the basis of our dataset, we estimate all the relevant physical parameters of SN 2012aw with our radiation-hydrodynamics code: envelope mass $M_{env} \sim 20 M_\odot$, progenitor radius $R \sim 3 \times 10^{13}$ cm ($ \sim 430 R_\odot$), explosion energy $E \sim 1.5$ foe, and initial $^{56}$Ni mass $\sim 0.06$ $M_\odot$. These mass and radius values are reasonably well supported by independent evolutionary models of the progenitor, and may suggest a progenitor mass higher than the observational limit of $16.5 \pm 1.5 M_\odot$ of the Type IIP events. 
\end{abstract}

%% Keywords should appear after the \end{abstract} command. The uncommented %%
%example has been keyed in ApJ style. See the instructions to authors %% for the
%journal to which you are submitting your paper to determine %% what keyword
%punctuation is appropriate.


%% From the front matter, we move on to the body of the paper. %% In the first
%two sections, notice the use of the natbib \citep %% and \citet commands to
%identify citations.  The citations are %% tied to the reference list via
%symbolic KEYs. The KEY corresponds %% to the KEY in the \bibitem in the
%reference list below. We have %% chosen the first three characters of the first
%author's name plus %% the last two numeral of the year of publication as our KEY %for %% each reference.

%% Authors who wish to have the most important objects in their paper %% linked in the electronic edition to a data center may do so by tagging %% their objects with \objectname{} or \object{}.  Each macro takes the %% object name as its required argument. The optional, square-bracket  %% argument should be used in cases where the data center identification %% differs from what is to be printed in the paper.  The text appearing  %% in curly braces is what will appear in print in the published paper.  %% If the object name is recognized by the data centers, it will be linked %% in the electronic edition to the object data available at the data centers   %% %% Note that for sources with brackets in their names, e.g. [WEG2004] 14h-090, %% the brackets must be escaped with backslashes when used in the first %% square-bracket argument, for instance, \object[\[WEG2004\] 14h-090]{90}). %%  Otherwise, LaTeX will issue an error. 

\section{Introduction}\label{intro}

Type II supernova (SN) events are the product of the collapse of a moderately massive progenitor, with an initial mass between $8 M_\odot$ (e.g. \citealt{pumo09}) and $30$ $M_\odot $ (e.g. \citealt{walmswell12}). According to the classical classification scheme (see \citealt{filippenko97} for a review) their spectra show prominent Balmer lines, which means that at the time of the explosion they have still retained their hydrogen-rich envelope. ``Plateau" Type II SNe (Type IIP) show a nearly constant luminosity for $ \sim 80-120$ days \citep{barbon79}. The plateau is an optically thick phase, in which the release of the thermal energy deposited by the shock wave on the expanding ejecta is driven by the hydrogen recombination front, which gradually recedes in mass (e.g. \citealt{kasen09}, \citealt{pumo11}). When the recombination front reaches the base of the hydrogen envelope, the light curve sharply drops by several magnitudes in $\sim 30$ days (e.g. \citealt{kasen09}; \citealt{olivares10}). This transition phase is followed by the linear ``radioactive tail'', powered by the decay of $^{56}$Co to $^{56}$Fe, which depends on the amount of $^{56}$Ni synthesized in the explosion (e.g. \citealt{weaver80}). In a few cases, the progenitors have been identified in high-resolution archival images and found to be to red supergiants (RSGs) of initial masses between $\sim 8 M_\odot$ and $\sim 17 M_\odot$. Available data show an apparent lack of high-mass progenitors, and this fact has been dubbed as the ``RSG problem'' \citep{smartt09}. \citet{walmswell12} suggested that the dust produced in the RSG wind could increase the line of sight extinction, with the net effect of underestimating the luminosity and, as a consequence, the mass of the progenitor. However, \citet{kochanek12} pointed out that all work to date, including that of \citet{walmswell12} has incorrectly used interstellar extinction laws rather than a consistent physical treatment of circumstellar extinction, which may lead to overestimate the effect of extinction. Finally, we note that there is evidence that a minor fraction of Type II SNe results from the explosion of blue supergiant stars, the best example being SN 1987A \citep{arnett89}. These SNe show a significant variety in the explosion parameters, but they generally display a Type IIP behaviour. \citet{smartt_etal_09} and \citet{pastorello12} have suggested that less than  $3-5\%$ of all Type II SNe are 1987A-like events.

The interest in Type IIP SNe is twofold. Firstly, observations show that
Type IIP SNe are the most common explosions in the nearby Universe
(e.g. \citealt{cappellaro99}; \citealt{li11}). This means that, given their
observed mass range, they can be used to trace the cosmic star formation history
up to $z \sim 0.6$ (see \citealt{botticella12}; \citealt{dahlen12}). Secondly,
it has been suggested that they can be used as cosmological distance indicators
(see \citealt{hamuy02}; \citealt{nugent06}; \citealt{poznanski09};
\citealt{olivares10}).

Despite their frequency and importance, only a fraction of Type IIP SNe has
been extensively monitored, photometrically and spectroscopically from the epoch of explosion through the late nebular phase. This type of extensive and extended monitoring is only viable for the closest events (typically closer than $10-15$ Mpc), as spectroscopic observations become difficult even with $10$m-class telescopes, beyond $300$ days. Examples with Type IIP SNe with this sort of coverage are SN 1999em \citep{elmhamdi03}, SN 1999gi \citep{leonard02}, SN 2004et \citep{maguire10}, SN 2005cs \citep{pastorello09}, SN 2009md \citep{fraser11}, SN 2012A \citep{tomasella13}.

Therefore, the occurrence of a nearby Type IIP SN offers us a unique
opportunity to collect very high quality photometric, spectroscopic and
polarimetric data from early stages up to the nebular phase. Through the
analysis of pre-explosion images we also have the possibility to compare the
progenitor parameters estimated with hydrodynamical explosion codes with the predictions of evolutionary models.

SN 2012aw was discovered by \citet{fagotti12} in the spiral galaxy M95 (NGC
3351), at the coordinates $\alpha_{2000}=10^{\rm h}43^{\rm m}53^{\rm s}.76$,
$\delta_{2000}=+11^{\rm o}40'17''.9$ on 2012 March $16.86$ UT. The magnitude at the discovery epoch was $R \sim 15$ mag and steeply rising ($R \sim 13$ mag, by J. Skvarc on March $17.90$ UT). The latest pre-discovery image was on March $15.86$ UT \citep{poznanski12atel}. These data allow us to constrain the explosion epoch to March $16.0 \pm 0.8$ UT, corresponding to the Julian Day (JD) 2,456,002.5 \citep{fraser12}. In the following, we will refer to this epoch as day $0$. The designation SN 2012aw was assigned after an early spectrum taken by \citet{munari12cbet} on 2012 March $17.77$ UT that showed  a very hot continuum without obvious absorption or emission features, and subsequently spectroscopic confirmations independently obtained by \citet{itoh12} and by \citet{siviero12} that showed a clear H$_\alpha$  P Cygni profile, indicating a velocity of the ejecta of about $15000$ km s$^{-1}$ \citep{siviero12}. 

SN 2012aw was also observed in the X-rays with \textit{Swift} \citep{immler12} between 2012 March 19.7 and March 22.2 UT at a luminosity $L_X = 9.2 \pm 2.5 \times 10^{38}$ erg s$^{-1}$, and at the radio frequency of $20.8$ GHz on March $24.25$ UT \citep{stockdale12} at a flux density of $0.160 \pm 0.025$ mJy. A subsequent radio observation on March 30.1 UT at the frequency of $21.2$ GHz revealed a flux density of $0.315 \pm 0.018  $ mJy \citep{yadav12}, thus confirming a radio variability. Finally, spectropolarimetric observations with VLT+FORS2 suggested a significant intrinsic continuum polarization at early phases, a possible signature of a substantial asymmetry in the early ejecta \citep{leonard12}.

A candidate progenitor was promptly identified as a RSG in archival Hubble
Space Telescope data by \citet{eliasrosa_atel12} and by \citet{fraser_atel12}. Detailed pre-SN multi-band photometry was carried out on space (HST WFPC2 $F814W$) and ground based (VLT+ISAAC, NTT+SOFI) archival images by \citet{fraser12}. Adopting a solar metallicity, they estimated a luminosity in the range $10^5 - 10^{5.6} L_\odot$ and an effective temperature between $3300$ and $4500$ K, and a progenitor radius larger than $500$ $R_\odot$. Their comparison with stellar evolutionary tracks pointed toward a progenitor with an initial mass between $14$ and $26$ $M_\odot$. We note that the uncertainties in the \citet{fraser12} parameters are mostly due to the line of sight extinction estimate, which they estimated to be larger than $E(B-V)=0.4$ mag at the $ 2 \sigma$ level and larger than $E(B-V) = 0.8$ mag at the $1 \sigma$ level. \citet{vandyk12} conducted a similar analysis, where they carefully discussed the infrared photometric calibration and the subtle effects due to the progenitor \textit{pre-explosion} reddening (which they estimated as $E(B-V)=0.71$ mag) and the variability of the RSG. They found the spectral energy distribution (SED) to be consistent with an effective temperature of $3600$ K, a luminosity $L \sim 10^{5.21} L_\odot$, a radius $R=1040$ $R_\odot$ and an initial mass between $15$ and $20$ $M_\odot$. After interpolating  their adopted tracks (taken from \citealt{ekstrom12}), they finally constrained the progenitor initial mass to be $\sim 17-18$ $M_\odot$, which is at the upper end of the initial masses for the Type IIP SNe progenitors detected to date, as suggested by \citet{smartt_etal_09}. Subsequently, \citet{kochanek12} suggested that the \citet{fraser12} and the \citet{vandyk12} progenitor luminosity (and mass) values may have been overestimated, since they adopted for the reddening the classical absorption-to-reddening ratio $R_V=3.1$, which is appropriate for a standard dust composition \citep{cardelli89}. \citet{kochanek12} pointed out that a massive RSG produces mostly silicates, for which a ratio of $R_V=2$ is more appropriate. Moreover, visual extinction may be overestimated, since the contribution of the scattered light in the interstellar extinction budget is neglected. In turn, they suggested a progenitor luminosity between $L=10^{4.8} L_\odot$ and $L=10^{5.0}L_\odot$ and a mass $M < 15$ $M_\odot$.

Accurate $BVRI$ light curves of SN 2012aw were published by \citet{munari13}, who carefully discussed the problems related to the homogenization of photometric measurements obtained at different telescopes, producing an optimal light curve by means of their ``lightcurve merging method''. Moreover, extensive photometric and spectroscopic observations were presented by \citet{bose13}, covering a period from $4$ to $270$ days after explosion. \citet{bose13} measured the photospheric velocity, the temperature and the $^{56}$Ni mass of SN 2012aw; they estimated the explosion energy and the mass of the progenitor star by comparing their data with existing simulations.

In this paper we present the results of our observational campaign, which include unpublished near-infrared data. We used our data for \textit{new} hydrodynamical simulations to estimate the relevant physical parameters. The same approach has been used for other two Type IIP SNe, namely SN 2012A \citep{tomasella13} and SN 2012ec (Barbarino et al. 2014, in prep.), thus providing a homogeneous analysis that can be used for consistent comparisons.

The paper is organized as follows: in Section \ref{m95}, we list the relevant properties of the host galaxy M95; in Section \ref{reddening} we discuss the reddening estimate, both Galactic and host; in Section \ref{phot}, we present our photometric dataset and analyze the photometric time evolution; in Section \ref{spec}, we present the spectroscopic observations and discuss the time evolution of the spectral features; in Section \ref{phys} we discuss the physical parameters obtained from the photometric and spectroscopic data: the bolometric light curve, from which we give an estimate of the $^{56}$Ni mass, the expansion velocity of the ejecta, and and SED evolution. In Section \ref{model}, we present the results of our hydrodynamical modelling, computed to match the observational parameters of SN 2012aw. Conclusions are presented in Section \ref{conclusions}.

\section{The host galaxy M95}\label{m95} 

M95 (NGC 3351, $\alpha_{2000}=10^{\rm h}43^{\rm m}57^{\rm s}.7$,
$\delta_{2000}= 11^{\rm o}42'12''.7$) is a face-on SBb(r)II spiral galaxy
\citep{sandage87} belonging to the Leo\,I Group. The total $V$-band magnitude is $M_V= -20.61 \pm 0.09$ mag and the total baryonic mass has been estimated as $M_{tot} = (3.57 \pm 0.30) \times 10^{10} M_\odot$ \citep{gurovich10}. The distance to M95 has been estimated with Cepheids and the tip of the red giant branch (TRGB). A range of distances have been reported during the years, but the latest estimates are comfortably converging: the HST Key Project gave a Cepheids-based distance of $(m-M)_0 = 30.00 \pm 0.09$ mag \citep{freedman01}, in excellent agreement with the TRGB-based distance of $(m-M)_0=29.92 \pm 0.05$ mag \citep{rizzi07}. This agreement is particularly striking, since it is based on two truly independent distance indicators, as Cepheids are young Population I stars, while the TRGB is a feature of the old Population II. A similar distance modulus was also obtained on the basis of the planetary nebulae luminosity function ($(m-M)_0= 30.0 \pm 0.16$ mag, \citealt{ciardullo02}). In the following, we will adopt as a distance modulus $(m-M)_0=29.96 \pm 0.04$ mag, which is the average of the Cepheids- and TRGB-based distances. M95 is known to host a central massive black hole (e.g. \citealt{beifiori09}), and its bulge shows intense star forming activity (e.g. \citealt{hagele07}). SN 2012aw is located in a southern outer arm, $60''$ west and $115''$ south of the
center of M95. The metallicity at the SN position can be approximated as
solar-like \citep{fraser12}. To our knowledge, no SN events were recorded in
M95 before SN 2012aw. Lastly, we note that the redshift of M95, as measured
from the H I 21 cm line, is $z=0.002598 \pm 0.000002$ \citep{springbob05}: we have adopted this value to redshift correct our spectra.

\section{Reddening}\label{reddening}   

In order to evaluate the physical parameters of the SN, photometric and
spectroscopic data have to be corrected for both the Galactic and the host galaxy reddening, and for the distance. The Galactic reddening was estimated using the \citet{schlegel98} maps, yielding $E(B-V)=0.028$ mag. We note that the new calibration of the dust maps, provided by \citet{schlafly11}, gives $E(B-V)=0.024$ mag. In the following discussion, we will adopt $E(B-V)=0.028$ mag for the Galactic reddening. 

The host galaxy reddening was estimated on the basis of the Na~ID equivalent width (EW) extracted from a SARG high-resolution spectrum. We measured EW(D2 $\lambda 5891) = 286 \pm 17$ m\AA ~\ and EW(D1 $\lambda 5897) = 240 \pm 16$ m\AA, corresponding to a column density of $log(NaI)=12.80 \pm 0.14$. As a first attempt, we used a classical (but still widely adopted in the literature, see for example \citealt{liszt14}) route to the reddening estimate: following \citet{ferlet85}, the Na~I column density value translates into $log(H)= 21.05 \pm 0.14$ and, according to \citet{bohlin78}, into a colour excess of $E(B-V)_{host}=0.19 ^{\pm 0.15}_{\pm 0.09}$ mag. The quoted uncertainty takes into account the $30\%$ uncertainty of the \citet{bohlin78} calibration only. This transforms into a relatively high host absorption of $A(B)_{host}=0.79 ^{\pm 0.62} _{\pm 0.37}$ mag, if a Galactic $R_V=3.1$ total-to-selective absorption ratio \citep{cardelli89} is assumed. We note that this large (but rather uncertain) value is in agreement, with the $E(B-V) = 0.15$ mag upper limit given by \citet{bose13}, on the basis of a blackbody fit to the early observed fluxes. Interestingly, by adopting the calibration given by \citet{turatto03} with the EW measured on the low-resolution spectra of \citet{bose13}, we get $E(B-V)_{host}=0.16$ mag. This reddening value is also in good agreement with the \citet{munari97} calibration (their Table 2), which suggests a reddening in the range $E(B-V)_{host} = 0.10 -0.12$ mag.

As an independent check we used the ``color-method'' \citep{olivares10}. This method relies on the assumption that, at the end of the plateau, the intrinsic $(V-I)$ colour is constant, and a possible colour-excess is only due to the host galaxy reddening (after correcting for the Galactic reddening). According to their eq. (7)

\begin{eqnarray} A_V(V-I)=2.518 [(V-I)-0.656] \\ 
\sigma(A_V)=2.518\sqrt{\sigma_{(V-I)}+0.053^2+0.059^2}
 \end{eqnarray}

and following the prescriptions described in their paper, we adopted in the
above formulae the $(V-I)$ colour at day $\sim 94$, corrected for the
foreground extinction, which is roughly $15$ days before the end of the
plateau. We derive $A(V)_{host} = (0.44 \pm 0.10)$ mag, which corresponds to $E(B-V)=0.14 \pm 0.03$ mag \citep{cardelli89}, in agreement with the other quoted estimates. 

It is interesting to note that our EW(Na~ID) measurements are, within the uncertainties, in excellent agreement with those obtained by \citet{vandyk12}, of $EW(D2) = 269 \pm 14$ \AA\  and $EW(D1) = 231 \pm 11$ \AA. \citet{vandyk12} derived a significantly lower reddening, of $E(B-V)_{host} = 0.055 \pm 0.014$ mag, by adopting the precise \citet{poznanski12} calibration. Consistently, \citet{bose13} obtained with the same method $E(B-V)_{host} = 0.041 \pm 0.011$ mag. These values are lower than those based on the other quoted methods, that point toward a reddening of $E(B-V) \sim 0.14$ mag. Interestingly, the latter value is consistent with the new $N(HI)/E(B-V)$ calibration provided by \citet{liszt14}, which gives $E(B-V)=0.13$ mag (for the sake of completeness, this calibration is referred to the relationship between the reddening and the atomic hydrogen column density only). However, both the \citet{bohlin78} and the \citet{liszt14} calibrations need an intermediate step to transform the Na~I column density into H column density, which adds its own uncertainty to the final estimate.

Therefore, we decided to follow our referee's suggestion to adopt the direct calibration of the reddening from the Na~I column density provided by \citet{poznanski12}, from which we get $E(B-V)_{host} = 0.058 \pm 0.016$ mag. This value translates into $A(B)_{host}=0.24 \pm 0.07$ mag. For the following discussion we will adopt a total extinction of $A(B)_{tot} = 0.36 \pm 0.07$ mag.

\section[]{Photometry}\label{phot} 
\subsection{Data} 

An intensive campaign of optical and near-infrared (NIR) observations of SN
2012aw was promptly started after its discovery (2012, March 17, day $0$), and lasted until the end of the plateau and the beginning of the radioactive tail phase (2012, July 21, day $130$), when the SN went into conjunction with the Sun. Two additional epochs were collected on 2012, December 26, and on 2013, February 11 (day $286$ and day $333$, respectively), well into the nebular phase.

Optical $UBVRI$ Johnson-Cousins images were collected with: the 67/92 cm Asiago Schmidt Telescope (Italy), equipped with a SBIG STL-11000M CCD camera (13 epochs); the array of $0.41$ m Panchromatic Robotic Optical Monitoring and Polarimetry Telescopes (PROMPT, Chile), equipped with Apogee U47p
cameras, which employ the E2V CCDs (33 epochs); the 2.2m telescope at the
Calar Alto Observatory (Spain), equipped with the CAFOS Focal Reducer and Faint Object Spectrograph instrument (2 epochs); the 1.82m Copernico telescope at Cima Ekar (Italy), equipped with the AFOSC Asiago Faint Object Spectrograph and Camera (2 epochs); the ESO NTT telescope (Chile), equipped with the EFOSC2 ESO Faint Object Spectrograph and Camera (2 epochs); the 4.2m William Herschel Telescope (WHT, Canary Islands, Spain), equipped with the ACAM Auxiliary Port Camera (2 epochs); and the 2.5m Nordic Optical Telescope (Canary Islands, Spain), equipped with the ALFOSC Andalucia Faint Object Spectrograph and Camera (3 epochs). Two early epochs, collected during the rise of the light curve and discussed in \citet{munari13}, have been included in our analysis for a better sampling of these phases.

Optical $ugriz$ Sloan data were collected with: the PROMPT Telescopes (21
epochs); the 2.0m Liverpool Telescope (LT, Canary Islands, Spain), equipped
with the RATCam optical CCD camera (11 epochs); and the 2.0m Faulkes Telescope North (Hawaii, USA), equipped with the FI CCD486 CCD detector (4 epochs).

NIR $JHK$ data were obtained with: the 0.6m Rapid Eye Mount (REM) Telescope
(Chile), equipped with the REMIR infrared camera (11 epochs); the 1.52m Carlos Sanchez Telescope (TCS, Canary Islands, Spain), equipped with the CAIN infrared camera (8 epochs); and the 3.58m Telescopio Nazionale Galileo (G, Canary Islands, Spain), equipped with the NICS Near Infrared Camera Spectrometer (1 epoch). 

Summarizing, our photometry densely covers the photospheric phase in the
$UBVRI$ and in $ugriz$ photometric systems, with $58$ epochs ranging from day $1.9$ to day $130$, and with $30$ epochs from day $3.1$ to day $114$,
respectively. Moreover, two additional epochs have been collected in $UBVRI$, during the nebular phase. Our NIR data are the only currently available in the literature for SN 2012aw, and they cover $17$ epochs from day $7.6$ to day $94$.

Data were pre-reduced by the instruments pipelines, when available, or with
standard procedures (bias and flat-field corrections, trimming; plus background subtraction for the NIR data) in the \texttt{IRAF} \footnote{IRAF is distributed by the National Optical Astronomical Observatory, which is operated by the  Association of Universities for Research in Astronomy, Inc., under cooperative agreement with the National Science Foundation.} environment. In a few cases, in which the sky background subtraction was not satisfactory, some NIR images were pre-reduced by means of an \texttt{IRAF}-based custom pipeline, which adopts for the background subtraction a two-step technique based on a preliminary guess of the sky background and on a careful masking of unwanted sources in the sky images, by means of the \texttt{XDIMSUM} \texttt{IRAF} package \citep{coppola11}.

Photometric measurements were carried out with the QUBA pipeline \citep{valenti11}, which performs PSF photometry on the SN and on selected field
stars. Johnson-Cousins $UBVRI$ magnitudes of the reference stars were calibrated by averaging the photometric sequence published in \citet{henden12} and our measurements obtained with the 67/92cm Asiago Schmidt Telescope; Sloan $ugriz$ reference star magnitudes were calibrated using images taken at the LT, during selected photometric nights. We did not transform the $ugriz$ dataset into the $UBVRI$ system, because the current state-of-the-art transformations \citep{jordi06}, which are appropriate for normal field stars, may not be accurate for SNe whose SED is strongly dominated by intense absorptions and emissions, which significantly alter the blackbody energy distribution.\footnote{The transformations between these two photometric systems may lead to systematic errors in the $u-g$ colour even for normal field stars, as the $u-g$ colour is particularly sensitive to temperature, surface gravity, and metallicity (e.g. \citealt{lenz98}).} Four reference stars in the $UBVRI$ system (namely, IDs $1, 2, 3$, and $7$) are in common with \citet{bose13}: the differences in the photometry are $-0.020 \pm 0.052$ mag, $-0.007 \pm 0.037$ mag, $0.012 \pm 0.035$ mag, $-0.002 \pm 0.059$ mag , and $-0.001\pm0.023$ mag in the $U$, $B$,$V$, $R$ and $I$ bands, respectively. Reference stars $1$ and $2$ also have Sloan Digital Sky Survey Data Release 9 (SDSS DR9, \citealt{ahn12}
measurements): the differences are $0.026 \pm 0.045$ mag, $0.028 \pm 0.024$ mag, $-0.008 \pm 0.012$ mag, $0.010 \pm 0.008$ mag, and $0.022 \pm 0.002$ mag, in the $u$, $g$, $r$, $i$ and $z$ bands, respectively. We point out that our adopted reference stars showed no clear signs of variability.

NIR data were calibrated by reference to four well measured Two Micron All Sky Survey (2MASS, \citealt{skrutskie06}) reference stars. We did not correct for the colour terms, since they are generally very small in the NIR bands (e.g. \citealt{carpenter01}) and the uncertainties of the photometric measurements were significantly larger than those related to neglecting the colour terms. Because of the small field of view, only one reference star was available in TCS images, and it was not possible to produce an accurate PSF model. We therefore adopted aperture photometry. However, we explicitly note that the SN is located far from the host galaxy's inner regions, and we do not expect a significant contamination of the background by the host galaxy. 

\begin{figure}  
\includegraphics[width=16.3cm]{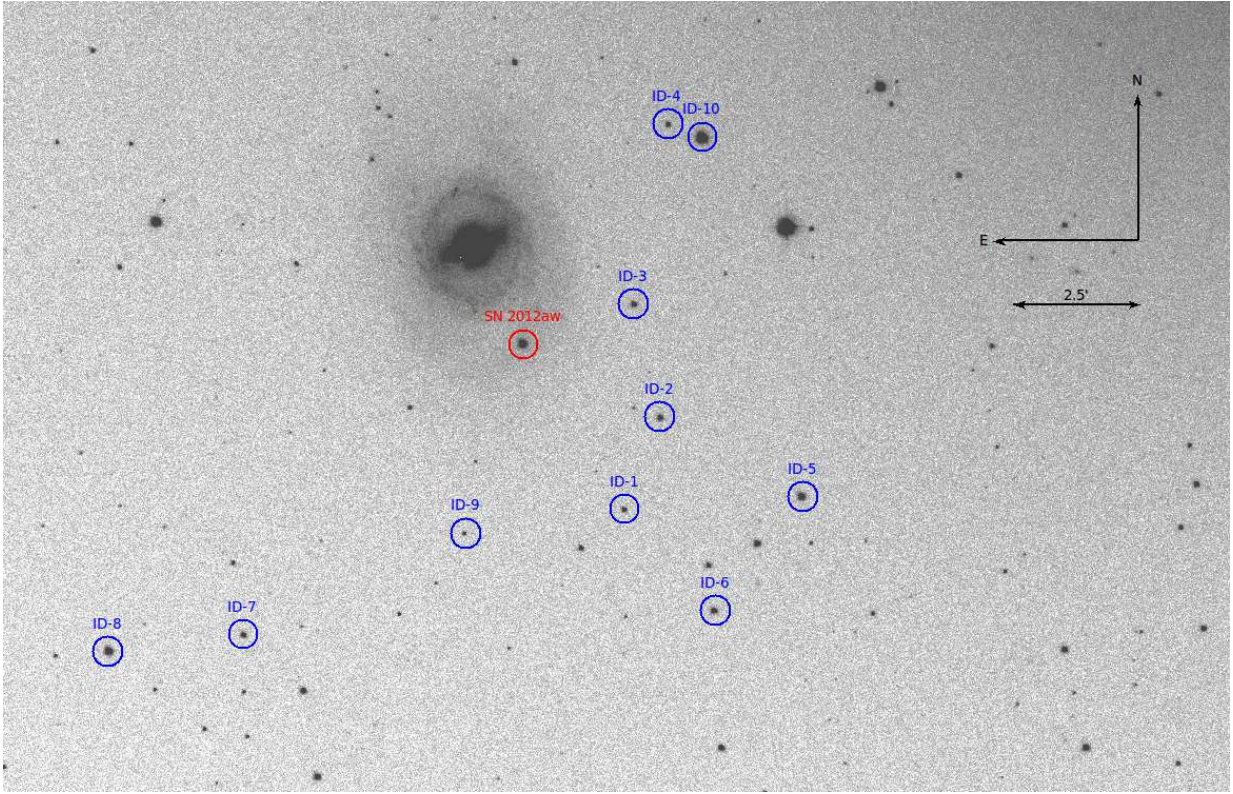} 
\caption{Finding chart of SN 2012aw and the reference stars. $V$-band image collected at the 67/92cm Asiago Schmidt Telescope on 2012, March 20. The area shown here is approximately $24 \times 15$ arcmin$^2$.} 
\label{sn2012_fc}
\end{figure}

Table \ref{map_cat} lists the positions and the photometric properties of the adopted reference stars, while a map of SN 2012aw and of the reference stars is shown in Figure \ref{sn2012_fc}. The photometry of SN 2012aw is reported in Tables \ref{log_ubvri}, \ref{log_ugriz} and \ref{log_jhk} for the $UBVRI$, $ugriz$, and $JHK$ systems, respectively. Reported photometric uncertainties are computed using the photometric errors and the uncertainties in the calibrations. When multiple exposures were available in the same night for the same filter, the adopted error was the rms of the measured magnitudes.

\begin{table} 
\begin{minipage}{130mm}  
\caption{Positions and photometry of the selected reference stars. $UBVRI$ and $ugriz$ magnitudes have been calibrated with Landolt fields on photometric nights; $JHK$ magnitudes have been taken from the 2MASS catalogue. Star IDs are the same for the three systems.} 
\label{map_cat} %\vspace{0.5 cm}  
\scriptsize
\begin{tabular}{cccccccc}  
\hline 
\hline 
Star ID & $\alpha_{J2000.0}$ &$\delta_{J2000.0}$ & {\it U} & {\it B} & {\it V} & {\it R} & {\it I} \\ 
&    (deg)         &     (deg)          &  (mag)  &  (mag)  &  (mag)  &   (mag) & mag 
\\ 
\hline 
1      &   $160.94117$     &    $11.617182$     
&                      & $16.384 \pm 0.018$      &   $15.620\pm0.006 $  &
$15.076\pm 0.009$  &  $14.694\pm 0.013$\\ 
2      &   $160.92930$     &   
$11.647304$      &  $15.729\pm0.008$    & $15.613 \pm 0.012$      &   $14.853\pm
0.020 $  &	$14.416\pm 0.020$  &  $14.018\pm 0.020$\\ 
3      &  
$160.93780$     &    $11.684113$      &  $15.221\pm0.009$    &
$15.351\pm0.040$      & $14.972\pm0.028$  & $14.706\pm0.026$  &
$14.450\pm0.012$\\ 
4      & $160.92599$      &  $ 11.743191$     
& 	     	        & $17.116\pm0.012$      & $15.821\pm0.005$ &
$14.952\pm0.006$   & $14.104\pm0.030$\\ 
5      & $160.88154$      & $
11.620989$        &                       & $14.992\pm0.026$      &
$13.932\pm0.038$ & $ 13.248\pm0.028$  & $12.717\pm0.034$\\ 
6      &
$160.91103$      & $ 11.583979 $     &                        &
$15.551\pm0.018$      & $14.669\pm0.020$ & $ 14.145\pm0.012$  &
$13.670\pm0.002$\\ 
7      & $161.06876$      & $ 11.576971$      
&                        & $15.706\pm0.009$      & $14.873\pm0.034$ & $
14.334\pm0.019$  & $13.949\pm0.004$\\ 
8      & $161.11392$      & $ 11.571762
$      &                        & $14.249\pm0.022$      & $13.516\pm0.029$ & $
13.088\pm0.026$  & $12.718\pm0.025$\\ 
%\end{tabular} %\begin{tabular}{cccccccc}
\hline \hline Star ID & $\alpha_{J2000.0}$ &$\delta_{J2000.0}$ & {\it u} & {\it
g} & {\it r} & {\it i} & {\it z} \\ &     (deg)         &     (deg)          & 
(mag)  &  (mag)  &  (mag)  &   (mag) &  mag \\ \hline 1      &   $160.94117$    
&    $11.617182$      &                      & $15.967 \pm 0.024$      &  
$15.372\pm0.021$    & $15.168\pm 0.015$  &  $15.105\pm 0.018$\\ 2      &  
$160.92930$     &    $11.647304$      &  $16.612\pm0.044$    & $15.244 \pm
0.018$      &   $14.653\pm 0.016 $  & $14.433\pm 0.008$  &  $14.312\pm 0.012$\\
3      &   $160.93780$     &    $11.684113$      &  $16.092\pm0.029$    &
$15.108\pm0.018 $      &   $14.883\pm0.016$  & $14.823\pm0.012$  &
$14.830\pm0.021$\\ \end{tabular}

\begin{tabular}{cccccc} \hline \hline Star ID & $\alpha_{J2000.0}$
&$\delta_{J2000.0}$ & {\rm J} & {\rm H} & {\rm K} \\ &     (deg)         &    
(deg)          &  (mag)  &  (mag)  &  (mag) \\ \hline 2      &   $160.92930$    
&    $11.647304$      & $13.380\pm0.027$     & $13.025\pm0.026$      &  
$12.914\pm0.034 $\\ 4      &   $160.94117$     &    $11.617182$      &
$13.163\pm0.026$     & $12.476\pm0.024$      &   $12.347\pm0.031$ \\ 9      &  
$160.93780$     &    $11.684113$      & $12.816\pm0.024$     & $12.218\pm0.024
$      &   $12.001\pm0.021$\\ 10     &   $160.91448$     &    $11.738809$      &
$10.233\pm0.027$     & $9.741\pm0.026 $      &   $9.554\pm0.027$\\ \end{tabular}
\end{minipage} \end{table}

%\clearpage

\clearpage
\begin{table} \begin{minipage}{130mm}  
\caption{Log of $UBVRI$ photometric observations of SN 2012aw. See text for details on the individuals instruments.}
\label{log_ubvri} %\vspace{0.5 cm}  
\scriptsize 
\begin{tabular}{lcccccccc} 
\hline 
\hline 

Date & JD        &  Phase\footnote{JD - 2,450,002.5}   & {\it U} & {\it B} & {\it V} & {\it R} & {\it I} & Source\footnote{1=Munari; 2 = Asiago
Schmidt Telescope; 3 = CAFOS; 5 = AFOSC; 6 = EFOSC2; 7 = ALFOSC; 12 = PROMPT; 13
= ACAM.}\\   & (2400000+)& (days)   &  (mag)  &  (mag)  &  (mag)  &  (mag)  &
(mag)   &       \\ \hline 17/03/2012 & 56004.41 & 1.9    & $              $ &
$13.79 \pm 0.01$ & $  13.86\pm 0.01 $ & $  13.82\pm 0.01 $ & $ 13.72\pm 0.01 $ &
1   \\ 18/03/2012 & 56005.57 & 3.1    & $              $ & $13.52\pm 0.05$ & $ 
13.68\pm 0.05$ & $  13.53\pm 0.03$ & $ 13.53\pm 0.01$ & 12  \\ 19/03/2012 &
56006.71 & 4.2    & $              $ & $13.47\pm 0.12$ & $  13.59\pm 0.11$ & $ 
13.40\pm 0.09$ & $ 13.39\pm 0.07$ & 12  \\ 19/03/2012 & 56006.41 & 3.9    &
$              $ & $13.60 \pm 0.01$ & $  13.58 \pm 0.01 $ & $  13.43\pm 0.01 $ &
$ 13.31\pm 0.01 $ & 1   \\ 19/03/2012 & 56006.44 & 3.9    & $              $ &
$13.54\pm 0.08$ & $  13.56\pm 0.07$ & $  13.35\pm 0.05$ & $ 13.41\pm 0.05$ & 2  
\\ 20/03/2012 & 56007.57 & 5.1    & $              $ & $13.47\pm 0.11$ & $ 
13.52\pm 0.10$ & $  13.29\pm 0.08$ & $ 13.28\pm 0.06$ & 12  \\ 20/03/2012 &
56007.31 & 4.8    & $              $ & $13.53\pm 0.06$ & $  13.52\pm 0.06$ & $ 
13.38\pm 0.05$ & $ 13.39\pm 0.05$ & 2   \\ 21/03/2012 & 56008.57 & 6.1    &
$              $ & $13.38\pm 0.05$ & $  13.39\pm 0.05$ & $  13.22\pm 0.03$ & $
13.20\pm 0.04$ & 12  \\ 21/03/2012 & 56008.31 & 5.8    & $              $ &
$13.41\pm 0.05$ & $  13.44\pm 0.05$ & $  13.27\pm 0.06$ & $ 13.22\pm 0.05$ & 2  
\\ 22/03/2012 & 56009.58 & 7.1    & $              $ & $13.42\pm 0.02$ & $ 
13.38\pm 0.02$ & $  13.11\pm 0.04$ & $ 13.13\pm 0.03$ & 12  \\ 22/03/2012 &
56009.31 & 6.8    & $              $ & $13.42\pm 0.04$ & $  13.36\pm 0.04$ & $ 
13.19\pm 0.04$ & $ 13.16\pm 0.02$ & 2   \\ 23/03/2012 & 56010.54 & 8.0    &
$              $ & $13.36\pm 0.02$ & $  13.34\pm 0.02$ & $  13.11\pm 0.03$ & $
13.11\pm 0.02$ & 12  \\ 23/03/2012 & 56010.35 & 7.8    & $              $ &
$13.43\pm 0.02$ & $  13.34\pm 0.02$ & $  13.18\pm 0.03$ & $ 13.11\pm 0.01$ & 2  
\\ 23/03/2012 & 56010.35 & 7.8    & $12.50\pm0.04  $ & $13.35\pm 0.02$ & $ 
13.30\pm 0.02$ & $  13.10\pm 0.01$ & $ 13.07\pm 0.03$ & 3   \\ 23/03/2012 &
56010.36 & 7.9    & $              $ & $13.39\pm 0.03$ & $  13.30\pm 0.03$ & $ 
13.12\pm 0.01$ & $ 13.12\pm 0.01$ & 2   \\ 24/03/2012 & 56011.54 & 9.0    &
$              $ & $13.38\pm 0.03$ & $  13.28\pm 0.03$ & $  13.11\pm 0.02$ & $
13.07\pm 0.02$ & 12  \\ 24/03/2012 & 56011.36 & 8.9    & $12.55\pm0.08  $ &
$13.32\pm 0.03$ & $  13.29\pm 0.02$ & $  13.12\pm 0.02$ & $ 13.06\pm 0.03$ & 3  
\\ 26/03/2012 & 56013.36 & 10.9   & $12.74\pm0.07  $ & $13.43\pm 0.06$ &
$	 	      $ & $         	  $ & $	             $ & 5   \\
26/03/2012 & 56013.39 & 10.9   & $              $ & $             $ & $ 
13.33\pm 0.03$ & $  13.16\pm 0.04$ & $ 13.08\pm 0.03$ & 2   \\ 27/03/2012 &
56014.44 & 11.9   & $              $ & $13.43\pm 0.03$ & $  13.31\pm 0.03$ &
$         	  $ & $	             $ & 2   \\ 28/03/2012 & 56015.53 & 13.0   &
$              $ & $13.51\pm 0.04$ & $  13.35\pm 0.03$ & $  13.13\pm 0.02$ & $
13.07\pm 0.02$ & 12  \\ 28/03/2012 & 56015.39 & 12.9   & $12.84\pm0.06  $ &
$13.50\pm 0.02$ & $  13.35\pm 0.02$ & $  13.12\pm 0.02$ & $ 13.07\pm 0.05$ & 5  
\\ 29/03/2012 & 56016.51 & 14.0   & $              $ & $13.48\pm 0.02$ & $ 
13.35\pm 0.02$ & $  13.11\pm 0.02$ & $ 13.03\pm 0.04$ & 12  \\ 29/03/2012 &
56016.37 & 13.9   & $              $ & $13.46\pm 0.03$ & $  13.30\pm 0.03$ & $ 
13.12\pm 0.03$ & $ 13.01\pm 0.01$ & 2   \\ 30/03/2012 & 56017.57 & 15.1   &
$              $ & $13.61\pm 0.09$ & $  13.34\pm 0.08$ & $  13.14\pm 0.03$ & $
12.98\pm 0.03$ & 12  \\ 30/03/2012 & 56017.37 & 14.9   & $              $ &
$             $ & $	    	      $ & $  13.08\pm 0.02$ & $ 13.02\pm 0.03$ &
12  \\ 31/03/2012 & 56018.43 & 15.9   & $              $ & $13.53\pm 0.02$ & $ 
13.29\pm 0.02$ & $  13.13\pm 0.03$ & $ 12.98\pm 0.02$ & 2   \\ 02/04/2012 &
56020.32 & 17.8   & $              $ & $13.58\pm 0.07$ & $  13.34\pm 0.06$ & $ 
13.13\pm 0.05$ & $ 12.92\pm 0.03$ & 2   \\ 11/04/2012 & 56029.53 & 27.0   &
$              $ & $             $ & $  13.37\pm 0.06$ & $  13.06\pm 0.08$ & $
12.90\pm 0.03$ & 12  \\ 14/04/2012 & 56032.60 & 30.1   & $              $ &
$             $ & $  13.40\pm 0.01$ & $  13.05\pm 0.01$ & $ 12.90\pm 0.05$ & 12 
\\ 17/04/2012 & 56035.55 & 33.0   & $              $ & $             $ & $ 
13.44\pm 0.02$ & $  13.09\pm 0.01$ & $ 12.88\pm 0.01$ & 12  \\ 24/04/2012 &
56042.43 & 39.9   & $              $ & $14.41\pm 0.03$ & $  13.42\pm 0.02$ & $ 
13.13\pm 0.03$ & $ 12.86\pm 0.02$ & 2   \\ 25/04/2012 & 56043.40 & 40.9   &
$              $ & $14.41\pm 0.04$ & $  13.44\pm 0.04$ & $  13.08\pm 0.03$ & $
12.84\pm 0.04$ & 2   \\ 25/04/2012 & 56043.49 & 41.0   & $              $ &
$14.43\pm 0.01$ & $  13.45\pm 0.01$ & $  13.06\pm 0.03$ & $ 12.91\pm 0.04$ & 7  
\\ 30/04/2012 & 56048.55 & 46.0   & $15.43\pm0.02  $ & $14.45\pm 0.02$ & $ 
13.46\pm 0.02$ & $  13.07\pm 0.02$ & $ 12.84\pm 0.02$ & 6   \\ 02/05/2012 &
56049.94 & 47.4   & $              $ & $             $ & $  13.50\pm 0.02$ & $ 
13.05\pm 0.03$ & $ 12.80\pm 0.04$ & 12  \\ 03/05/2012 & 56050.57 & 48.1   &
$              $ & $14.54\pm 0.04$ & $  13.46\pm 0.04$ & $  13.07\pm 0.04$ & $
12.79\pm 0.03$ & 12  \\ 06/05/2012 & 56053.40 & 50.9   & $15.70\pm0.03  $ &
$14.72\pm 0.02$ & $  13.54\pm 0.02$ & $  13.07\pm 0.04$ & $ 12.82\pm 0.05$ & 13 
\\ 09/05/2012 & 56056.61 & 54.1   & $              $ & $             $ & $ 
13.53\pm 0.02$ & $  13.08\pm 0.03$ & $ 12.81\pm 0.02$ & 12  \\ 12/05/2012 &
56059.65 & 57.2   & $              $ & $14.71\pm 0.08$ & $  13.53\pm 0.08$ & $ 
13.10\pm 0.04$ & $ 12.80\pm 0.01$ & 12  \\ 21/05/2012 & 56069.55 & 67.0   &
$              $ & $15.10\pm 0.06$ & $  13.56\pm 0.05$ & $  13.02\pm 0.03$ & $
12.76\pm 0.03$ & 12  \\ 23/05/2012 & 56071.57 & 69.1   & $              $ &
$             $ & $  13.59\pm 0.03$ & $  13.02\pm 0.04$ & $ 12.90\pm 0.04$ & 12 
\\ 26/05/2012 & 56074.38 & 71.9   & $16.56\pm0.05  $ & $14.97\pm 0.01$ & $ 
13.60\pm 0.01$ & $  13.08\pm 0.02$ & $ 12.83\pm 0.02$ & 7   \\ 27/05/2012 &
56075.61 & 73.1   & $              $ & $             $ & $  13.59\pm 0.02$ & $ 
13.02\pm 0.03$ & $ 12.75\pm 0.03$ & 12  \\ 07/06/2012 & 56086.55 & 84.0   &
$              $ & $             $ & $  13.64\pm 0.01$ & $  13.11\pm 0.01$ &
$	             $ & 12  \\ 13/06/2012 & 56092.51 & 90.0   & $             
$ & $             $ & $  13.67\pm 0.03$ & $  13.11\pm 0.03$ & $	             $ &
12  \\ 17/06/2012 & 56096.41 & 93.9   & $17.17\pm0.06  $ & $15.19\pm 0.02$ & $ 
13.75\pm 0.02$ & $  13.17\pm 0.01$ & $ 12.88\pm 0.01$ & 7   \\ 24/06/2012 &
56103.53 & 101.0  & $              $ &	            & $  13.82\pm 0.04$ & $ 
13.18\pm 0.05$ & $ 12.88\pm 0.04$  & 12  \\ 26/06/2012 & 56105.40 & 102.9  &
$              $ & $15.45\pm 0.04$ & $  13.88\pm 0.02$ & $  13.21\pm 0.04$ & $
12.85\pm 0.06$ & 13  \\ 02/07/2012 & 56111.48 & 109.0  & $              $ &
$15.31\pm 0.12$ & $  13.90\pm 0.11$ & $  13.29\pm 0.05$ & $ 12.95\pm 0.03$ & 12 
\\ 06/07/2012 & 56115.49 & 113.0  & $              $ & $15.45\pm 0.04$ & $ 
14.01\pm 0.03$ & $  13.37\pm 0.05$ & $ 13.08\pm 0.06$ & 12  \\ 07/07/2012 &
56116.40 & 113.9  & $17.62\pm0.05  $ & $15.47\pm 0.03$ & $  14.03\pm 0.03$ & $ 
13.40\pm 0.02$ & $ 13.11\pm 0.02$ & 7   \\ 08/07/2012 & 56117.48 & 115.0  &
$              $ & $15.49\pm 0.11$ & $  14.02\pm 0.10$ & $  13.40\pm 0.02$ & $
13.12\pm 0.02$ & 12  \\ 09/07/2012 & 56118.49 & 116.0  & $              $ &
$15.52\pm 0.05$ & $  14.05\pm 0.05$ & $  13.37\pm 0.03$ & $ 13.08\pm 0.02$ & 12 
\\ 17/07/2012 & 56126.48 & 123.0  & $              $ & $15.87\pm 0.07$ & $ 
14.32\pm 0.03$ & $  13.63\pm 0.03$ & $ 13.28\pm 0.03$ & 12  \\ 19/07/2012 &
56128.48 & 126.0  & $              $ & $15.92\pm 0.12$ & $  14.46\pm 0.11$ &
$         	  $ & $	             $ & 12  \\ 20/07/2012 & 56129.48 & 127.0  &
$              $ &	              &  	    	& $  13.88\pm 0.04$ & $
13.57\pm 0.05$ & 12  \\ 23/07/2012 & 56132.47 & 130.0  & $              $
&	            & $  14.67\pm 0.01$ & $  13.88\pm 0.02$ &	               &
12  \\ 26/12/2013 & 56288.70 & 286.2  & $              $ & $18.55\pm 0.02$ & $ 
17.37\pm 0.02$ & $  16.36\pm 0.04$ & $ 15.90\pm 0.03$ &  7  \\ 11/02/2013 &
56335.63 & 333.1  & $20.34\pm0.10  $ & $18.98\pm 0.03$ & $  17.80\pm 0.02$ & $ 
16.85\pm 0.01$ & $ 16.32\pm 0.02$ & 13  \\

\hline 
\hline 
\end{tabular} 
\end{minipage}  
\end{table}

\begin{table} 
\begin{minipage}{130mm} 
\caption{Log of $ugriz$ photometric
observations of SN 2012aw. See text for details on the individual instruments.}
\label{log_ugriz}  %\vspace{0.5 cm}  
\scriptsize
\begin{tabular}{lcccccccc} \hline \hline Date & JD        &  Phase\footnote{JD -
2,450,002.5}   & {\it u} & {\it g} & {\it r} & {\it i} & {\it z} &
Source\footnote{2 = Asiago Schmidt Telescope; 4=RATCAM; 10 = Faulkes North; 12 =
PROMPT.}\\   & (2400000+)& (days)   &  (mag)  &  (mag)  &  (mag)  &  (mag)  &
(mag)   &       \\ \hline 18/03/2012 & 56005.57 & 3.1  & $ 	       $&
$13.57\pm 0.04 $ & $13.68\pm 0.03 $ & $13.87\pm 0.02 $ & $14.00\pm 0.02 $
&12\\   19/03/2012 & 56006.58 & 4.1  & $ 	       $& $13.46\pm 0.03 $ &
$13.57\pm 0.02 $ & $13.73\pm 0.01 $ & $13.84\pm 0.01 $ &12\\   20/03/2012 &
56007.57 & 5.1  & $ 	       $& $13.50\pm 0.04 $ & $13.46\pm 0.03 $ &
$13.62\pm 0.01 $ & $13.75\pm 0.02 $ &12\\   21/03/2012 & 56008.57 & 6.1  & $
13.34\pm 0.11 $& $13.40\pm 0.03 $ & $13.38\pm 0.02 $ & $13.54\pm 0.02 $ &
$13.67\pm 0.02 $ &12\\   22/03/2012 & 56009.58 & 7.1  & $ 	       $&
$13.38\pm 0.03 $ & $13.32\pm 0.02 $ & $13.49\pm 0.01 $ & $13.62\pm 0.02 $
&12\\   23/03/2012 & 56010.54 & 8.0  & $ 	       $& $13.33\pm 0.02 $ &
$13.31\pm 0.02 $ & $13.47\pm 0.01 $ & $13.58\pm 0.02 $ &12\\   23/03/2012 &
56010.36 & 7.9  & $ 13.29\pm 0.04 $& $13.26\pm 0.03 $ & $13.30\pm 0.02 $ &
$13.43\pm 0.01 $ & $13.52\pm 0.01 $ &2 \\   24/03/2012 & 56011.54 & 9.0  &
$ 	       $& $13.33\pm 0.03 $ & $13.28\pm 0.03 $ & $13.41\pm 0.01 $ &
$13.54\pm 0.02 $ &12\\   25/03/2012 & 56012.06 & 9.6  & $ 	       $&
$13.30\pm 0.12 $ & $13.24\pm 0.13 $ & $13.45\pm 0.04 $ & $13.54\pm 0.07 $
&10\\   26/03/2012 & 56013.49 & 11.0 & $ 13.55\pm 0.14 $& $13.27\pm 0.02 $ &
$13.26\pm 0.01 $ & $13.40\pm 0.02 $ & $13.49\pm 0.02 $ &4 \\   28/03/2012 &
56015.53 & 13.0 & $ 	       $& $13.38\pm 0.05 $ & $13.29\pm 0.05 $ &
$13.38\pm 0.01 $ & $13.47\pm 0.02 $ &12\\   29/03/2012 & 56016.51 & 14.0 &
$ 	       $& $13.37\pm 0.04 $ & $13.27\pm 0.03 $ & $13.40\pm 0.02 $ &
$13.50\pm 0.02 $ &12\\   30/03/2012 & 56017.37 & 14.9 & $ 	       $&
$13.34\pm 0.11 $ & $13.32\pm 0.10 $ & $13.37\pm 0.03 $ & $13.43\pm 0.02 $
&12\\   06/04/2012 & 56024.41 & 21.9 & $ 14.39\pm 0.06 $& $13.42\pm 0.02 $ &
$13.18\pm 0.02 $ & $13.26\pm 0.03 $ & $13.28\pm 0.01 $ &4 \\   11/04/2012 &
56029.53 & 27.0 & $ 	       $& $		 $ & $13.21\pm 0.01 $ &
$13.28\pm 0.01 $ & $13.34\pm 0.01 $ &12\\   14/04/2012 & 56032.60 & 30.1 &
$ 	       $& $		 $ & $13.24\pm 0.02 $ & $13.29\pm 0.02 $ &
$13.27\pm 0.03 $ &12\\   16/04/2012 & 56034.56 & 32.1 & $ 	       $&
$13.76\pm 0.04 $ & $13.21\pm 0.04 $ & $13.26\pm 0.01 $ & $		  $ &4
\\   17/04/2012 & 56035.55 & 33.0 & $ 	       $& $		 $ & $13.25\pm
0.02 $ & $13.31\pm 0.03 $ & $13.26\pm 0.01 $ &12\\   21/04/2012 & 56039.41 &
36.9 & $ 16.18\pm 0.08 $& $13.82\pm 0.01 $ & $13.25\pm 0.01 $ & $13.27\pm 0.02 $
& $13.24\pm 0.01 $ &4 \\   09/05/2012 & 56056.61 & 54.1 & $ 	       $&
$		 $ & $13.27\pm 0.02 $ & $13.23\pm 0.02 $ & $13.14\pm 0.02 $
&12\\   14/05/2012 & 56061.58 & 59.1 & $ 	       $& $14.11\pm 0.03 $ &
$13.26\pm 0.03 $ & $13.21\pm 0.02 $ & $13.12\pm 0.01 $ &12\\   21/05/2012 &
56069.55 & 67.0 & $ 	       $& $		 $ & $13.27\pm 0.02 $ &
$13.22\pm 0.02 $ & $13.04\pm 0.04 $ &12\\   23/05/2012 & 56071.57 & 69.1 &
$ 	       $& $		 $ & $13.30\pm 0.01 $ & $13.22\pm 0.01 $ &
$		  $ &12\\   26/05/2012 & 56074.43 & 71.9 & $ 17.67\pm 0.16 $&
$14.22\pm 0.02 $ & $13.27\pm 0.02 $ & $13.19\pm 0.01 $ & $13.10\pm 0.01 $ &4
\\   27/05/2012 & 56075.61 & 73.1 & $ 	       $& $		 $ & $13.27\pm
0.02 $ & $	       $ & $13.11\pm 0.02 $ &12\\   31/05/2012 & 56079.41 & 76.9
& $ 17.90\pm 0.12 $& $14.22\pm 0.03 $ & $13.30\pm 0.02 $ & $13.20\pm 0.02 $ &
$13.11\pm 0.02 $ &4 \\   01/06/2012 & 56080.41 & 77.9 & $ 17.84\pm 0.10 $&
$14.26\pm 0.03 $ & $13.27\pm 0.03 $ & $13.21\pm 0.02 $ & $13.09\pm 0.02 $ &4
\\   07/06/2012 & 56086.55 & 84.0 & $ 	       $& $		 $ & $13.29\pm
0.03 $ & $13.29\pm 0.03 $ & $13.21\pm 0.06 $ &12\\   24/06/2012 & 56103.53 &
101.0& $ 	       $& $		 $ & $13.42\pm 0.02 $ & $13.36\pm 0.02 $
& $13.23\pm 0.02 $ &12\\   07/07/2012 & 56116.48 & 114.0& $ 	       $&
$		 $ & $13.62\pm 0.02 $ & $13.54\pm 0.02 $ & $13.36\pm 0.01 $
&12\\  
\hline    
\end{tabular} 
\end{minipage}  
\end{table}

\begin{table} \begin{minipage}{130mm} \caption{Log of NIR observations of the SN
2012aw. See text for details on the individual instruments.}\label{log_jhk} 
%\vspace{0.5 cm}  
\scriptsize \begin{tabular}{lcccccc} \hline \hline Date &
JD        &  Phase\footnote{JD - 2,450,002.5}   & {\it J} & {\it H} & {\it K} &
Source\footnote{8 = NICS; 9= REM; 11 = TCS.}\\   & (2400000+)& (days)   & 
(mag)  &  (mag)  &  (mag)  &        \\ \hline 23/03/2012 &  56010.07 & 7.6  &
$13.00\pm 0.06 $ & $ 12.95\pm 0.06 $ & $ 12.66\pm 0.06 $ & 9  \\   24/03/2012 & 
56011.09 & 8.6  & $13.04\pm 0.04 $ & $ 12.87\pm 0.04 $ & $ 12.71\pm 0.07 $ & 9 
\\   25/03/2012 &  56012.12 & 9.6  & $12.90\pm 0.04 $ & $ 12.78\pm 0.04 $ & $
12.52\pm 0.04 $ & 9  \\   29/03/2012 &  56016.68 & 14.2 & $12.82\pm 0.10 $ & $
12.63\pm 0.07 $ & $ 12.45\pm 0.06 $ & 8  \\   01/04/2012 &  56019.07 & 16.6 &
$12.80\pm 0.04 $ & $ 12.62\pm 0.04 $ & $ 12.56\pm 0.04 $ & 9  \\   04/04/2012 & 
56022.08 & 19.6 & $12.74\pm 0.03 $ & $ 12.57\pm 0.03 $ & $ 12.42\pm 0.04 $ & 9 
\\   07/04/2012 &  56025.07 & 22.6 & $		$ & $ 12.55\pm 0.08 $ & $
12.35\pm 0.04 $ & 9  \\   13/04/2012 &  56031.37 & 28.9 & $12.56\pm 0.07 $ & $
12.39\pm 0.08 $ & $		$ & 11 \\   17/04/2012 &  56035.01 & 32.5 &
$12.54\pm 0.05 $ & $ 12.34\pm 0.05 $ & $ 12.26\pm 0.04 $ & 9  \\   22/04/2012 & 
56040.38 & 37.9 & $		$ & $		    $ & $ 11.96\pm 0.17 $ & 11
\\   24/04/2012 &  56042.12 & 39.6 & $12.49\pm 0.09 $ & $ 12.34\pm 0.09 $ & $
12.14\pm 0.09 $ & 9  \\   02/05/2012 &  56049.99 & 47.5 & $12.41\pm 0.04 $ & $
12.21\pm 0.04 $ & $ 12.08\pm 0.06 $ & 9  \\   04/05/2012 &  56052.42 & 49.9 &
$12.34\pm 0.04 $ & $ 12.03\pm 0.06 $ & $		$ & 11 \\   15/05/2012
&  56063.41 & 60.9 & $12.49\pm 0.04 $ & $ 12.74\pm 0.07 $ & $		$ & 11
\\   06/06/2012 &  56085.40 & 82.9 & $12.31\pm 0.02 $ & $ 12.18\pm 0.06 $ &
$		$ & 11 \\   10/06/2012 &  56088.41 & 85.9 & $12.34\pm 0.04 $ & $
12.12\pm 0.04 $ & $ 11.97\pm 0.11 $ & 11 \\   17/06/2012 &  56096.44 & 93.9 &
$12.42\pm 0.03 $ & $ 12.23\pm 0.06 $ & $ 12.04\pm 0.01 $ & 11 \\  

\hline    
\end{tabular} 
\end{minipage}  
\end{table}

\subsection{Time Evolution} \label{phot_evol}

\begin{figure}  
\includegraphics[width=16.3cm]{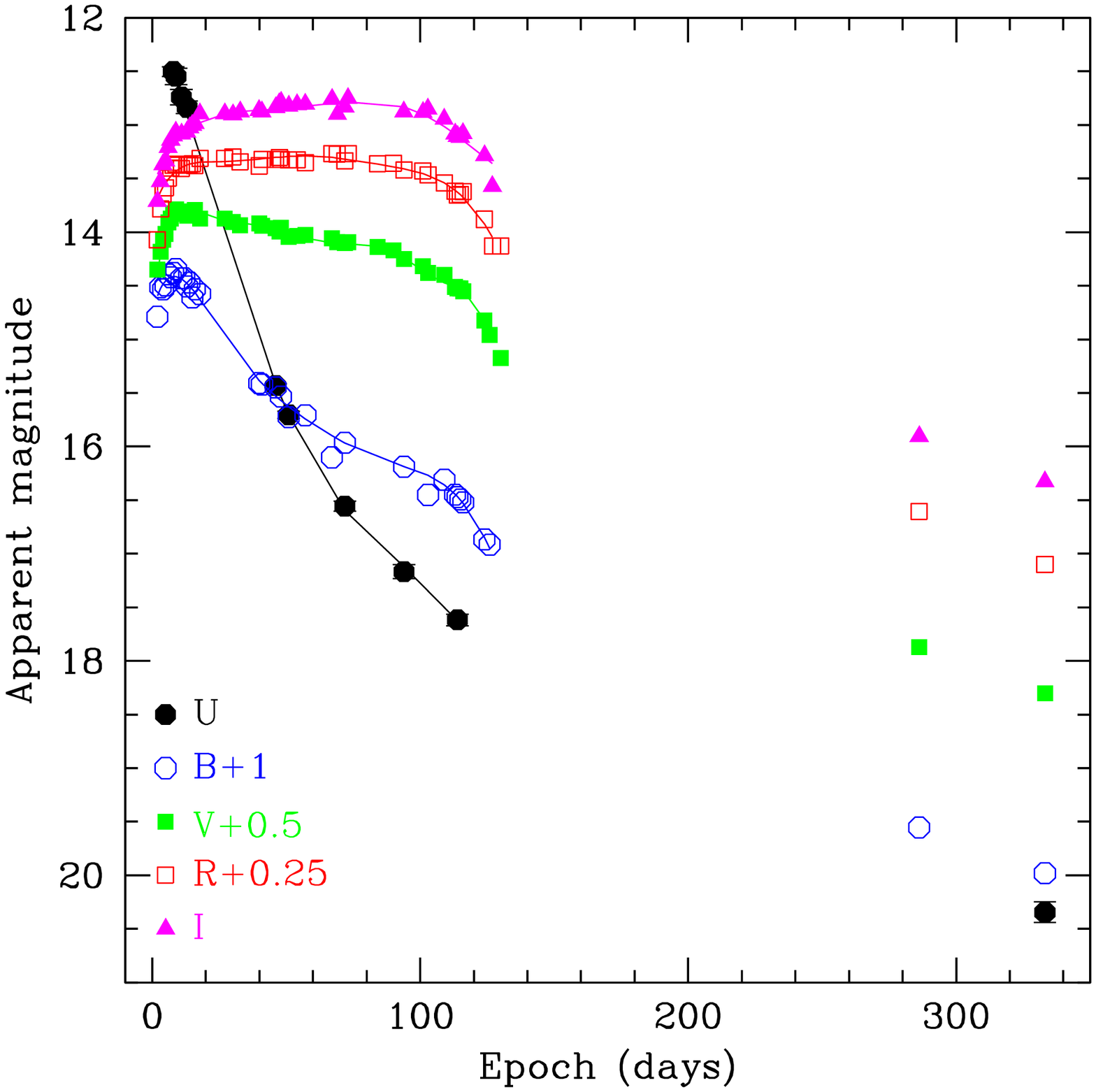}
\caption{Photometric evolution in the $UBVRI$ system. Individual light curves were shifted for clarity. Day $0$ corresponds to the adopted explosion epoch.} \label{optical_lc} 
\end{figure}

\begin{figure}  
\includegraphics[width=16.3cm]{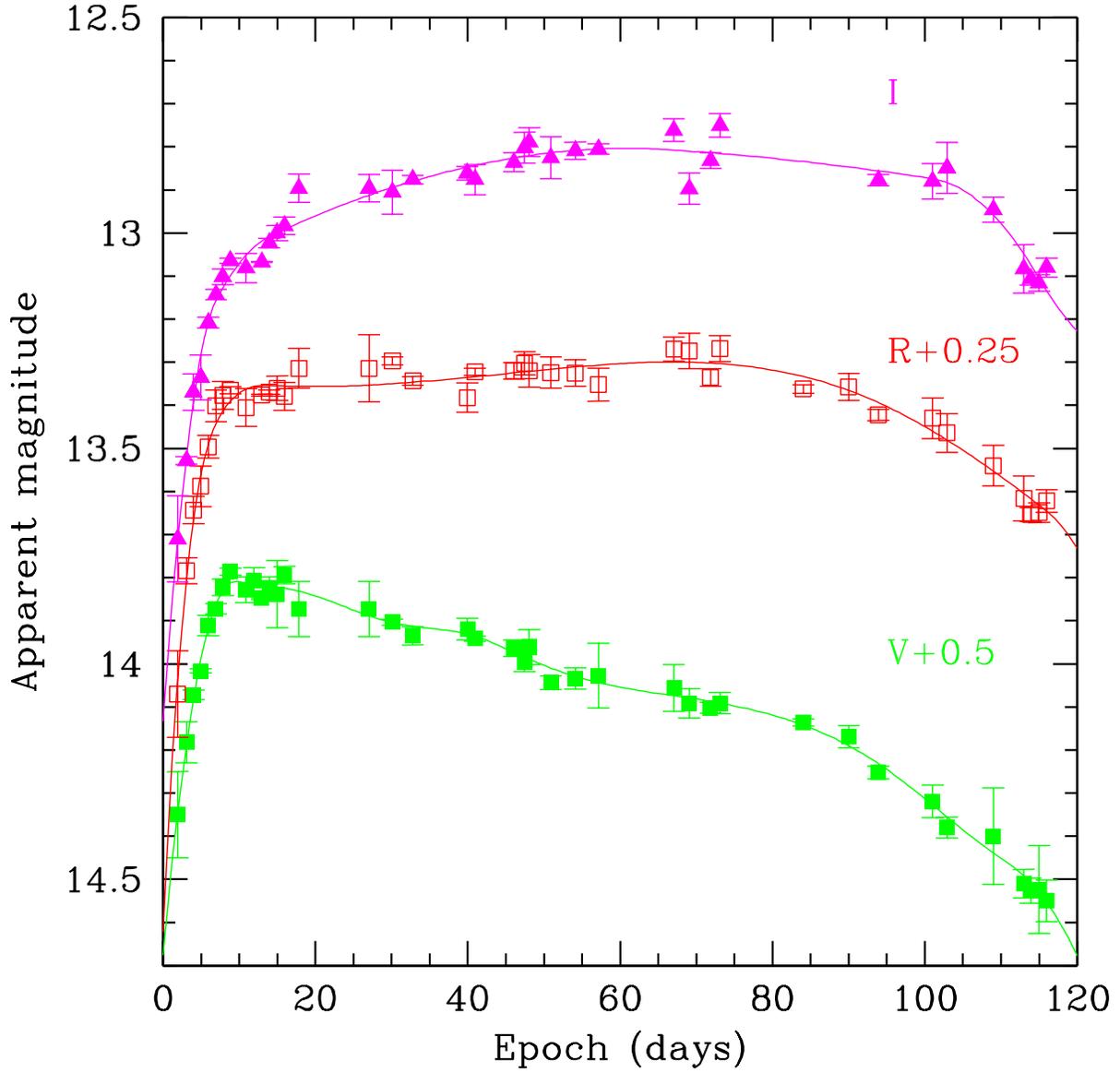} 
\caption{Photometric evolution in the $V$, $R$, and $I$ bands, between day $1$ and day $120$. Data points are interpolated with cubic splines, for visualization purposes.}
\label{vri_zoom} 
\end{figure}

\begin{figure}  
\includegraphics[width=16.3cm]{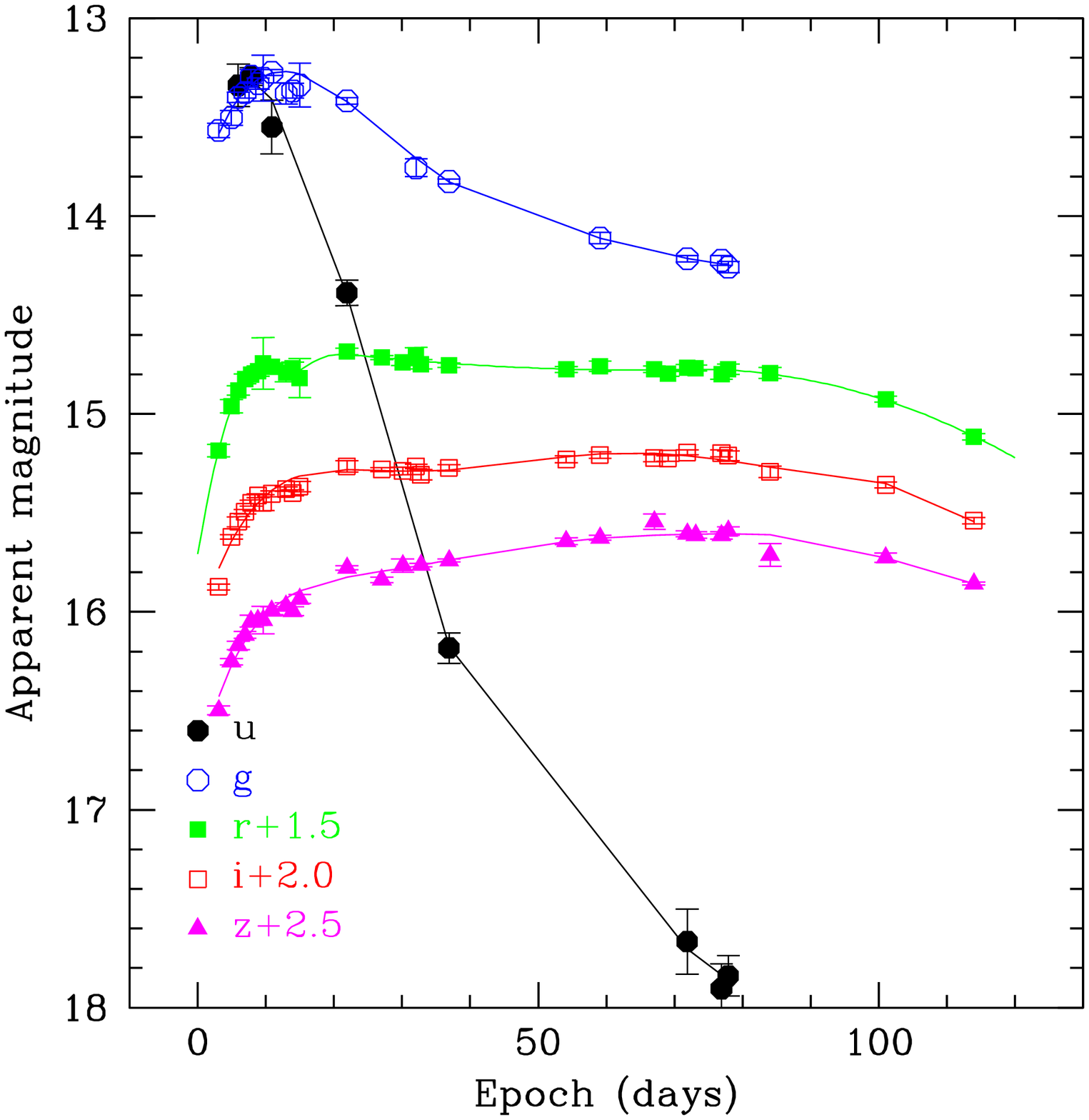}
\caption{Photometric evolution in the $ugriz$ system. Individual light curves were shifted for clarity.}  \label{sloan_lc} 
\end{figure}

\begin{figure}  
\includegraphics[width=16.3cm]{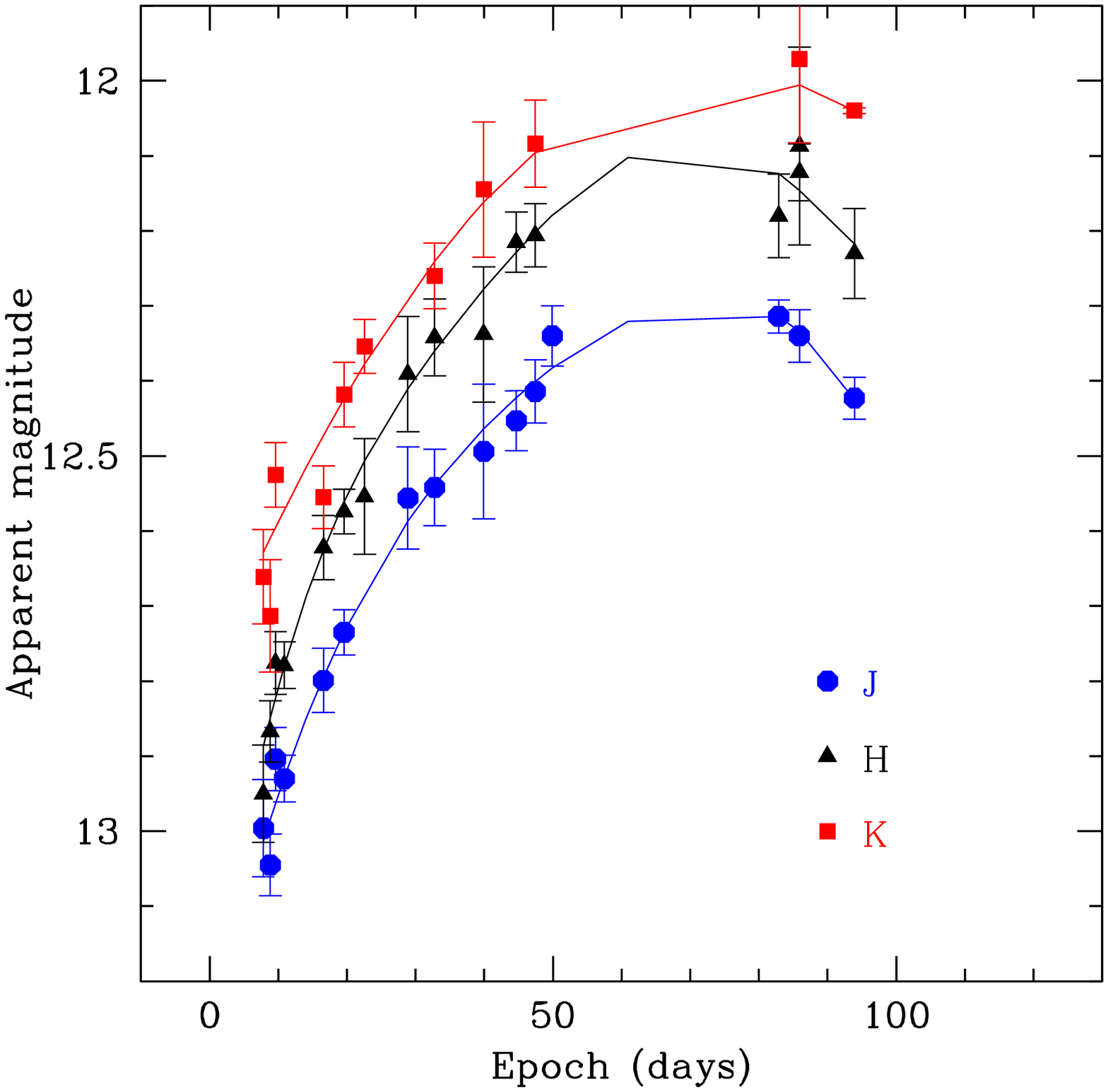}
\caption{Photometric evolution in the $JHK$ system.} 
\label{nir_lc} 
\end{figure}

We were able to follow the photospheric phase of SN 2012aw up to day $\sim
130$, observing the end of the plateau phase. Figures \ref{optical_lc},
\ref{sloan_lc}, and \ref{nir_lc} show the photometric evolution of SN 2012aw in the Johnson-Cousins, Sloan and NIR photometric systems, respectively. Figure (\ref{vri_zoom}) shows a close-up of the $V$, $R$, and $I$ light curves in the first $\sim 120$ days. Error bars are typically smaller than the symbol size, except for the NIR plot. Solid curves represent Chebyshev polynomials fitted to the observed data points, with the \texttt{CURFIT} \texttt{IRAF} task. The order of the fit was allowed to vary, to minimize the $\chi^2$. The rms was generally of the order of $\sim 0.03$ mag. In a few cases ($U$, $u$, and NIR bands) the sampling was poor and we adopted a cubic spline. The last two points, collected in the SN nebular phase, were not included in the fit. The plotted light curves show that the SN was discovered well before the $V$-band maximum, estimated from the fit at Julian Day $2,456,011.8 \pm 0.5$ (day $9.3 \pm 0.5$). A comparison of the early spectra of SN 2012aw (see Sect. \ref{spec_details}) with the collection of spectra available through the web tool \texttt{GELATO} \citep{harutyunyan08} independently confirms our estimate of the epoch of the explosion. The Johnson $U$ and $B$ light curves show a steady decline from day $2$ and day $\sim 7$ onwards, respectively, whereas $V$, $R$ and $I$ bands show the typical plateau behaviour of Type IIP events. The plateau lasts for $\sim 100$ days (also confirmed in \citealt{bose13}), followed by the drop to the radioactive tail. The Sloan photometry is consistent with such a behaviour. Unfortunately, we do not have convincing evidence of the minima in the $V$, $R$, and $I$ bands, claimed by \citet{bose13} at day $42$, $39$, and $31$, respectively. Our $V$ band photometry shows a quite constant decline from day $\sim 25$ to day $\sim 120$, i.e. to the end of the plateau phase; the $R$ band light curve suggests a sharp rise to the plateau phase, which in this band lasts from day $\sim 10$ to day $85$; the $I$ band light curve also reveals a sharp rise up to day $\sim 10$, followed by a slower rise to day $\sim 60$ and a stable plateau that lasts up to day $\sim 100$. Our Sloan $r$, $i$, and $z$ light curves behave consistently. Interestingly, we observe a possible small flattening in the $I$ band at day $\sim 10$, followed by a quite steep rise between day $\sim 12$ and day $17$, also visible in the Sloan $r$, $i$, and $z$ bands. Finally, the NIR $J$, $H$, $K$ photometry shows a steady brightening up to day $64$, with a behaviour similar to other Type IIP SNe (e.g. SN 2005cs, \citealt{pastorello09}). The apparent drop at the day $\sim 95$ could be an artifact, due the poor quality of the TCS data, where only one reference star was available.

\begin{figure}   
\includegraphics[width=8.3cm]{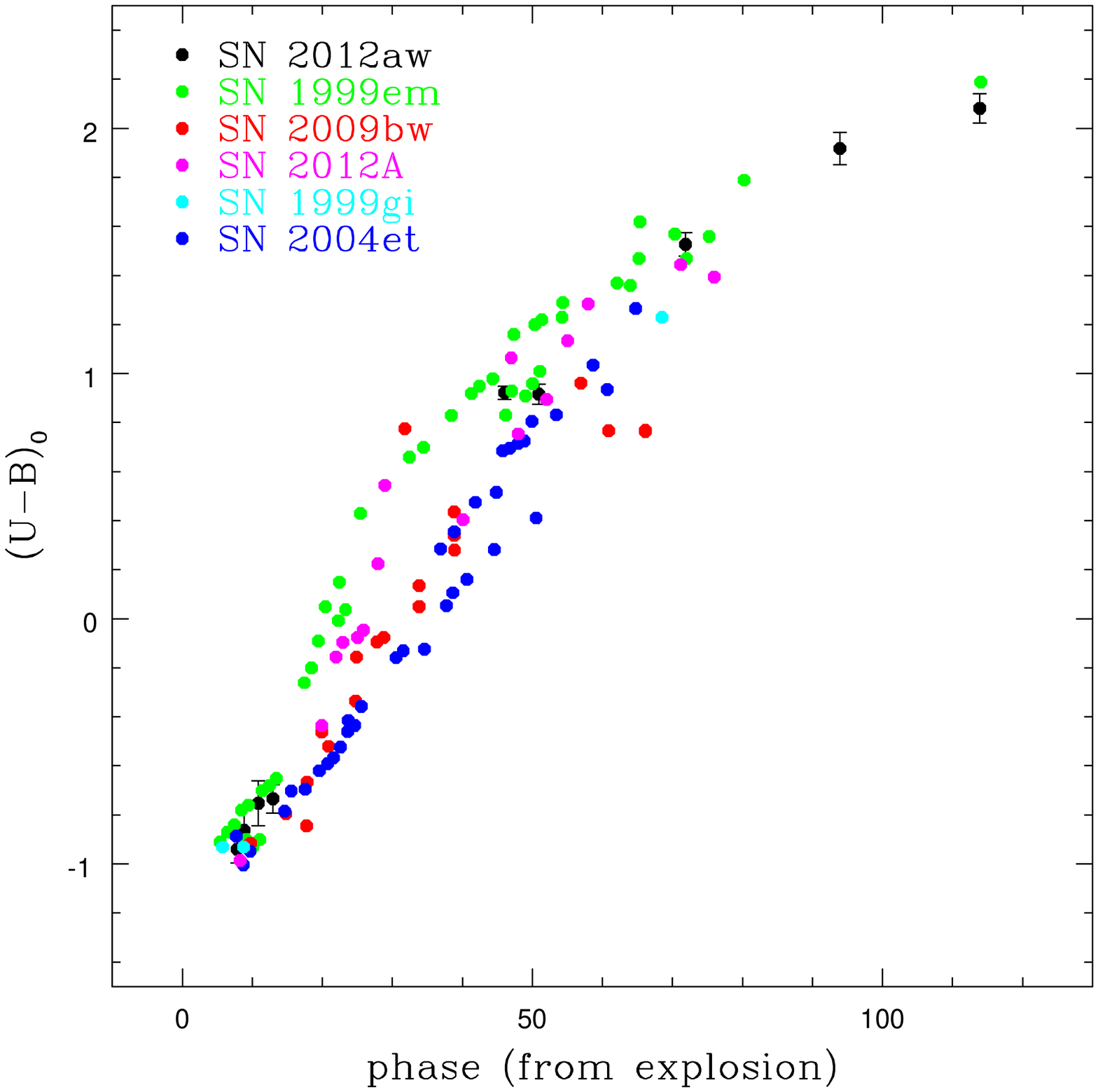} 
\includegraphics[width=8.3cm]{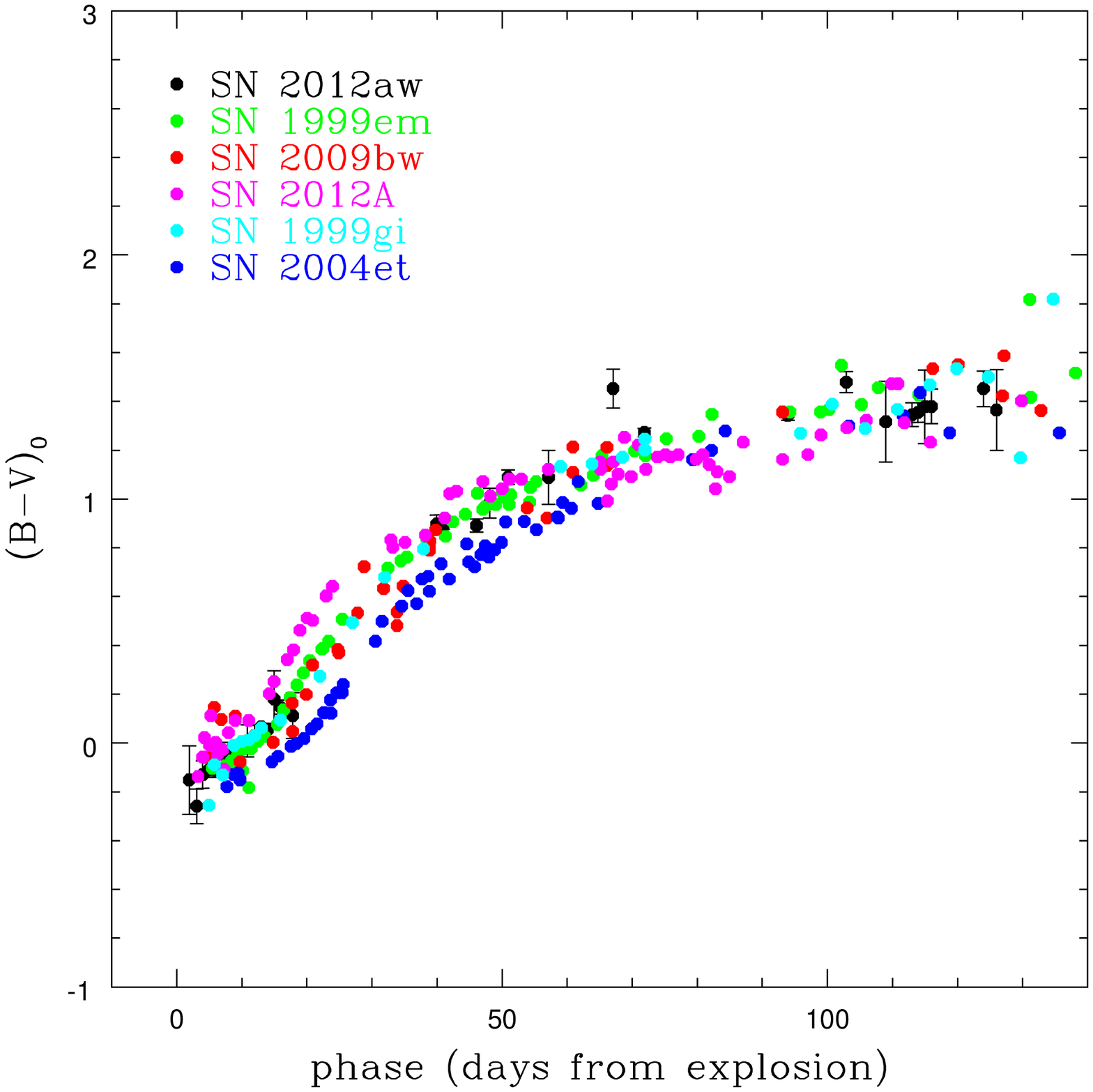} 
\includegraphics[width=8.3cm]{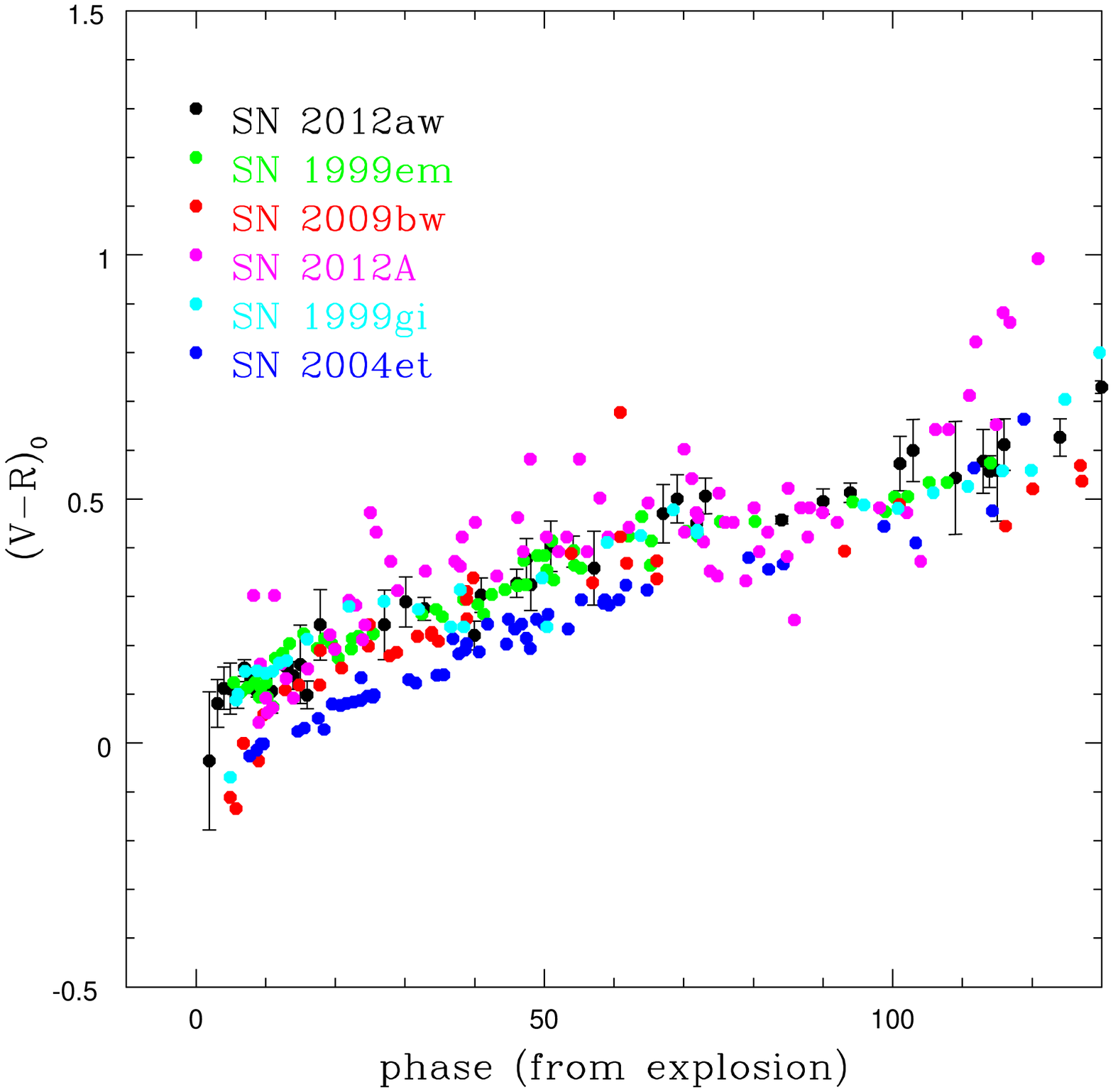}
\includegraphics[width=8.3cm]{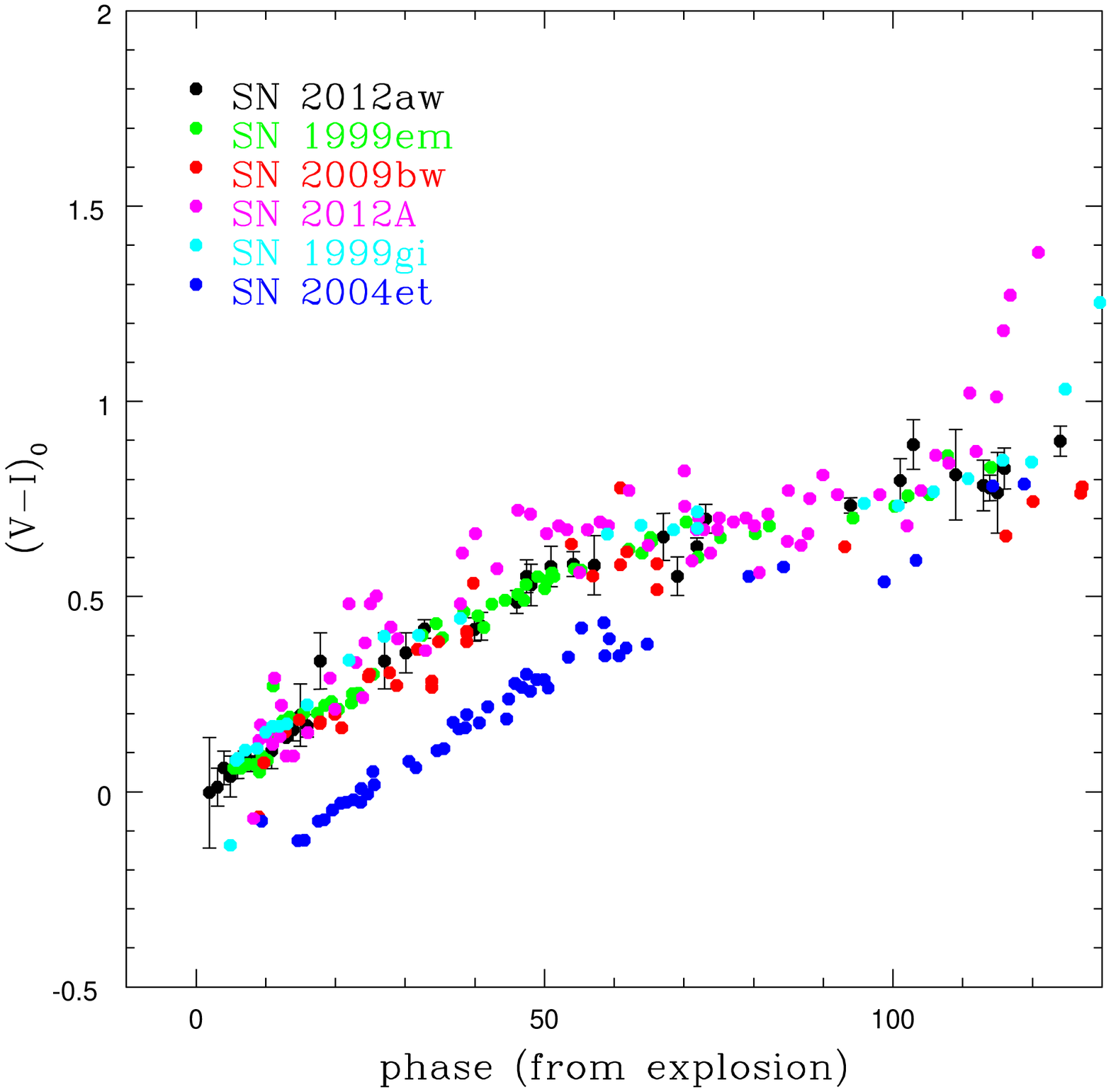}  
\caption{Dereddened colour evolution of SN 2012aw in the $UBVRI$ system, compared with other Type IIP SNe in the literature. The adopted extinction coefficients were taken from the papers quoted in the text.}  
\label{col_optical} 
\end{figure}

Figure \ref{col_optical} shows the $U-B$, $B-V$, $V-R$ and $V-I$ colour
evolution of SN 2012aw during the photospheric phase, compared to those of
other literature SNe. Colours of all SNe have been dereddened (see sec.\ref{reddening}), for a proper comparison. The colour evolution appears to be similar to that of other Type IIP SNe in the literature, namely SN 2012A \citep{tomasella13}, SN 1999em \citep{elmhamdi03}, SN 2009bw \citep{inserra12}, SN 1999gi \citep{leonard02}, and SN 2004et \citep{maguire10}. The plots show that SN 2012aw follows the typical evolution of Type IIP events, with a rapidly increasing $B-V$ colour for the first $40$ days, followed by a flattening of the curve.

\begin{figure}   
\includegraphics[width=8.3cm]{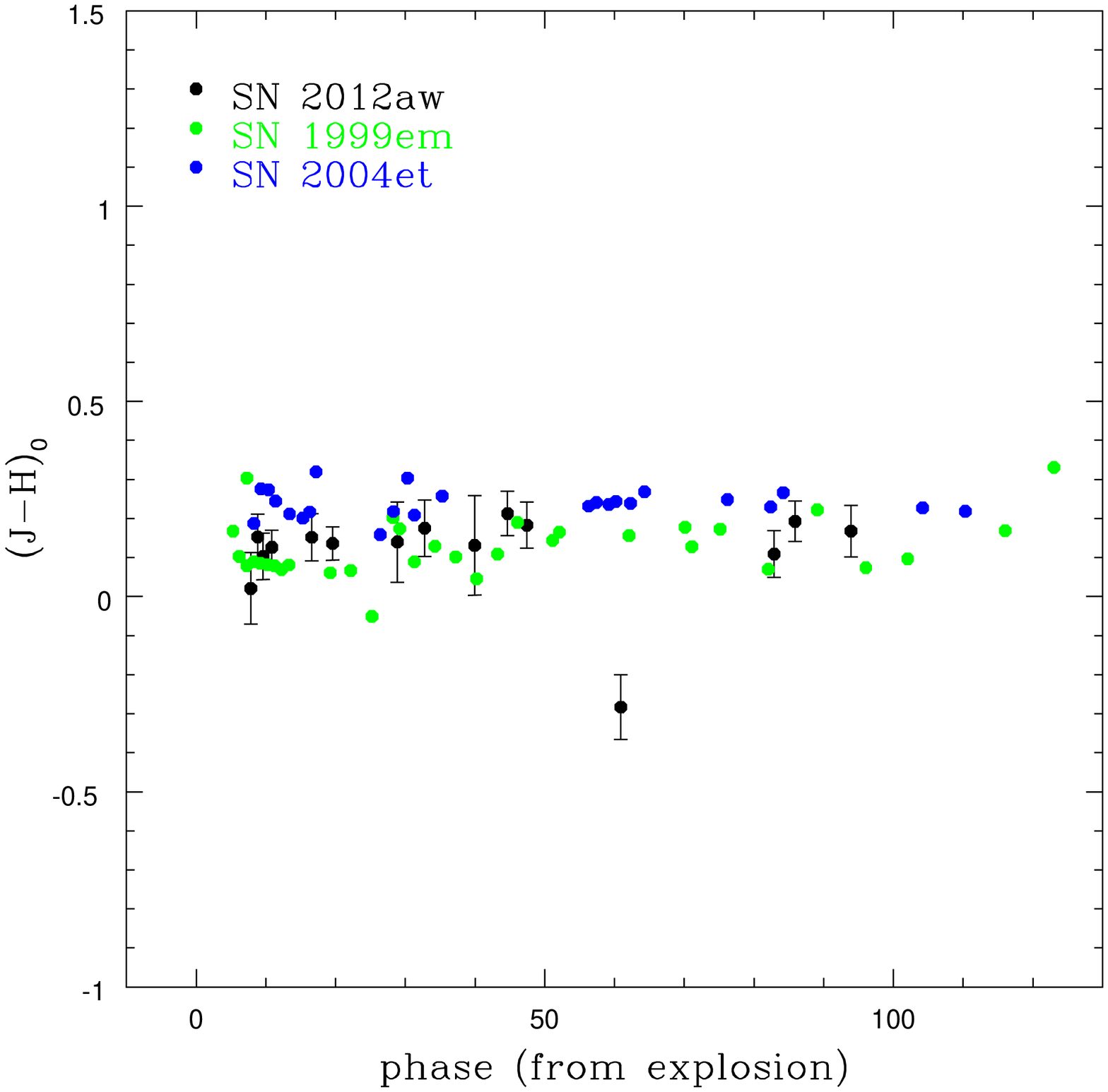} 
\includegraphics[width=8.3cm]{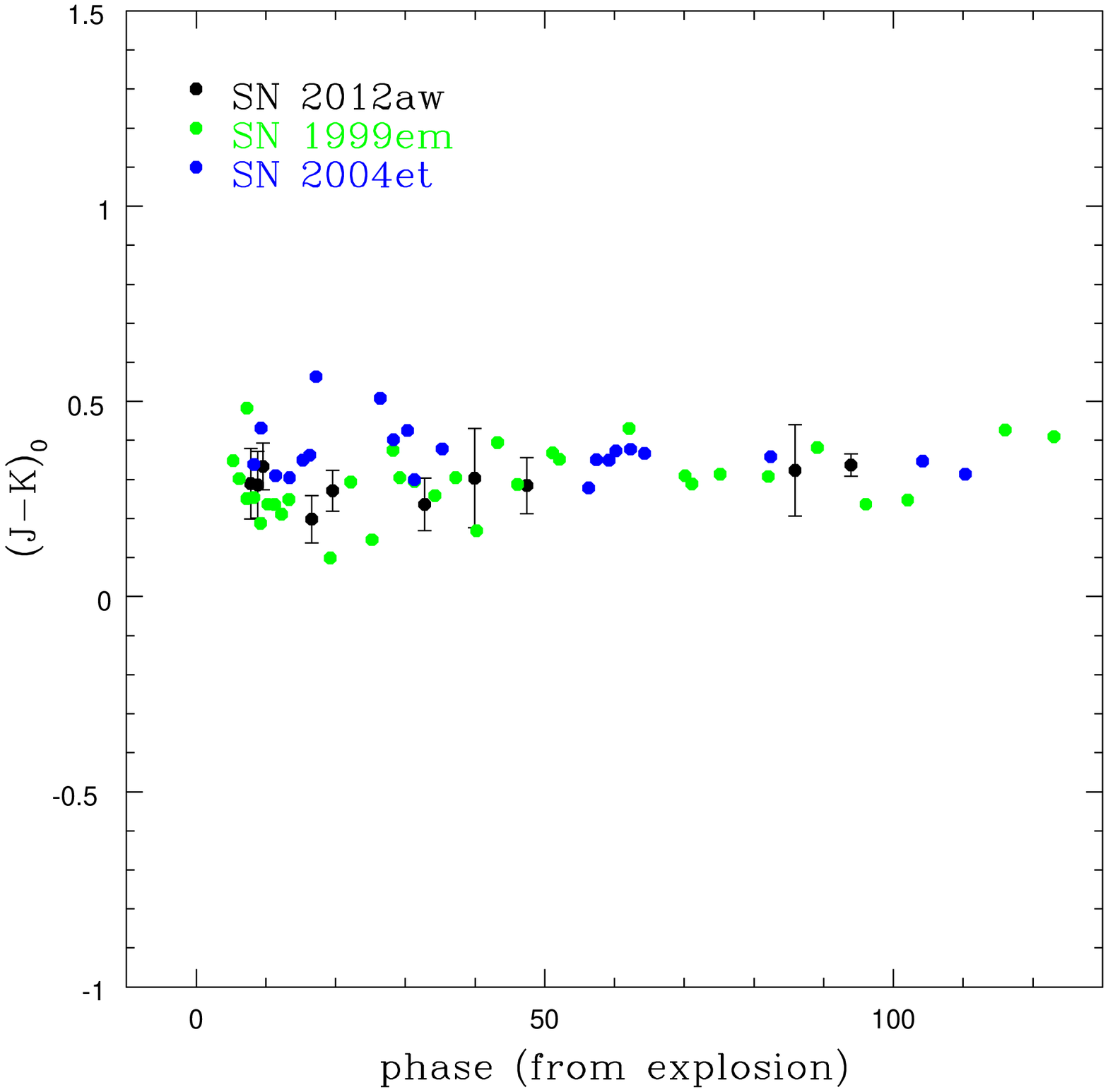} 
\caption{$(J-H)$ and $(J-K)$ colour evolution of SN 2012aw, compared with SN 1999em and SN 2004et. Individual colour curves have been dereddened according to the papers quoted in the text.} 
\label{col_nir} 
\end{figure}

Finally, Figure \ref{col_nir} depicts the time evolution of the intrinsic NIR colours $J-H$ and $J-K$. For sake of completeness, we also show the colour curves of SN 1999em \citep{elmhamdi03} and of SN 2004et \citep{maguire10}, for which we have a satisfactory time coverage in the NIR bands. Individual colour curves show a rather large scatter, likely due to the photometric errors, but overall the three SNe show a similar behaviour with a very small colour evolution during the monitored period.

%\begin{figure}  
%\includegraphics[width=16.3cm]{col_gr_rel3.ps}
%\caption{Dereddened $(g-r)$ colour curve of SN 2012aw, compared with SN 2009kf.}
%\label{col_sdss} 
%\end{figure}

\section{Spectroscopy}\label{spec}

\subsection{Spectroscopic observations and data reduction}  

Spectroscopic data were collected mostly during the first three months of
evolution.  We followed the spectroscopic evolution over $35$ epochs from day $1$ to day $94$, in a wavelength range from $3300$ to $25000$ \AA. Optical long-slit medium resolution spectra were collected with: the Boller \& Chivens spectrograph at the Asiago $1.22$m telescope ($3300-7800$ \AA, $12$ epochs); ALFOSC at the NOT $2.56$m ($3200 - 9100$ \AA, $5$ epochs); AFOSC at the Ekar $1.82$m ($3500-11000$ \AA, $4$ epochs); DOLORES at the TNG $3.58$m ($3000-10000$ \AA, $2$ epochs); EFOSC2 at the NTT ($3700-9300$, $2$ epochs); CAFOS at the CAHA $2.2$m ($3200 - 7000$ \AA, $1$ epoch); and ISIS at the WHT ($5400-9500$ \AA, $1$ epoch). Near-infrared low resolution spectra were obtained with: FIRE at the Magellan $6.5$m
telescope ($8000-25000$ \AA, $4$ epochs); and NICS at the TNG ($9000 - 25000$ \AA, $1$ epoch). High-resolution spectra were collected with SARG at the TNG ($4600-7900$\AA, $1$ epoch; and $5000 - 10100$ \AA, $1$ epoch), and with ISIS at the WHT ($3500-5200$ \AA, $1$ epoch). Table \ref{log_spectra} lists all the spectroscopic observations, with the instruments and the instrumental setups. 

FIRE (Folded-Port Infrared Echellette) spectra were reduced using a custom-developed IDL pipeline \citep{hsiao13}. All other spectra, were pre-reduced in a standard fashion (overscan and bias subtraction, trimming, flat-fielding) by using the tools available in \texttt{IRAF}. Wavelength calibration was carried out taking spectra of arc lamps with the same instrumental setup used for the science observations. Calibrated spectra were corrected for the heliocentric recessional velocity of the host galaxy. Flux calibration was performed through a comparison with selected spectrophotometric standard stars, obtained during the same nights as the scientific observations and with the same instrumental setup. Finally, the absolute flux calibration of the spectra was verified by comparing the integrated flux in the $UBVRI$ bands, measured using the \texttt{IRAF} package \texttt{CALCPHOT}, with the corresponding photometric measurements. When the spectra were collected on nights for which no photometry was available, a simple average of the adjacent photometric measurements was adopted. For spectra not bracketed by two consecutive photometric measurements, the polynomial light curve, discussed in the previous section, was used as a reference. Differences between the spectro-photometric and the photometric fluxes where corrected by multiplying and fitting the spectra with suitable coefficients. After the correction, the difference between the spectro-photometric and the photometric magnitudes were between $0.01$ and $0.05$ mag. The same procedure was adopted for the NICS near-infrared spectra. It is worth noting that \texttt{CALCPHOT} adopts the \citet{bessel88} NIR photometric system, while our photometry was calibrated
onto the 2MASS system. We therefore transformed the \texttt{CALCPHOT} synthetic photometry into the 2MASS system following \citet{carpenter01}. Finally, we corrected the spectra for the adopted reddening.

\subsection{Spectral Time Evolution} \label{spec_details}

\begin{figure}  
\includegraphics[width=16.3cm]{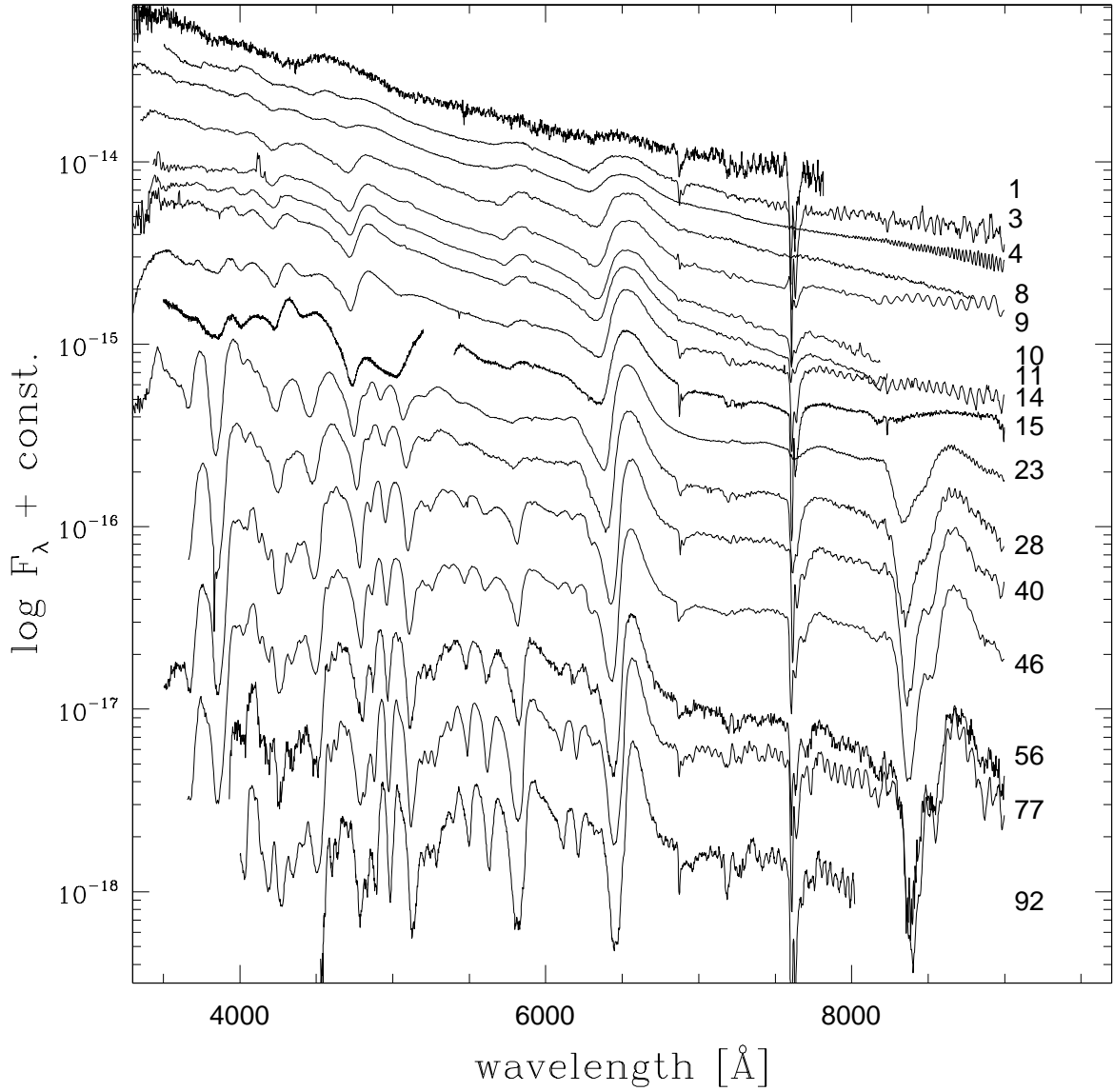}
\caption{Spectral time evolution of SN 2012aw. Individual spectra are scaled in flux by an arbitrary quantity, for clarity. Numbers on the left indicate the epoch from core-collapse.} 
\label{spectral_atlas} \end{figure}

\begin{figure}  
\includegraphics[width=16.3cm]{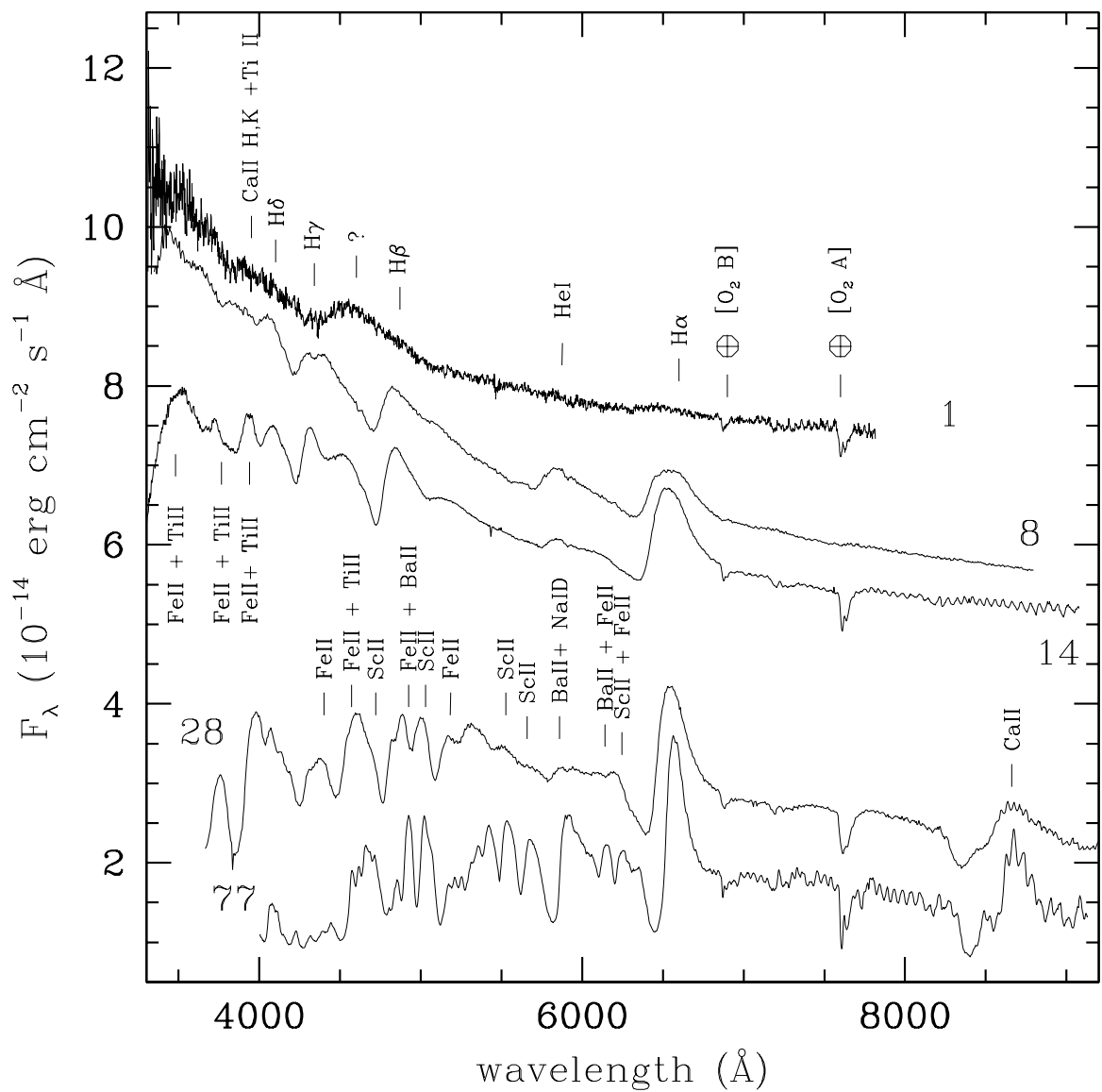} 
\caption{Line identification of the most prominent features of selected spectra of SN 2012aw. Individual spectra have been shifted in flux by an arbitrary quantity, for clarity. Numbers on the two sides show the epochs from the explosion.}
\label{lines} 
\end{figure}

Figure \ref{spectral_atlas} shows the optical spectral evolution of SN 2012aw, with the phases relative to the adopted explosion epoch, while a comprehensive atlas of the identified features is shown in Figure \ref{lines}, at relevant phases. The first spectrum, taken less than two days after the estimated explosion, exhibits an almost featureless hot continuum. Interestingly enough, a ``bump-shaped'' feature is clearly visible at about $4600$ \AA. This bump fades very quickly, and it is no longer visible at the epoch of $V$-band maximum (day $\sim 9$). A similar feature was also reported and discussed for SN 2009bw \citep{inserra12}. A possible identification is with a blend of highly ionized C and N features (also discussed for the Type IIn event SN 1998S, \citealt{fassia01}). The second spectrum, collected on day $\sim 3$, shows the emergence of the typical H$_\alpha$ line, as well as the He~I feature at $\sim 5876$ \AA. Initially, the H$_\alpha$ line shows a weak absorption component and a boxy emission. This feature may be the signature of a weak interaction with the circumstellar medium (see also SN 2007od, \citealt{inserra11}). This is also suggested by early radio observations (\citealt{stockdale12}; \citealt{yadav12}). The He~I feature is no longer visible after day $15$, while slightly blueward of He~I a possible blend  of the sodium doublet Na~ID ($ 5890, 5896$ \AA) with Ba~II appears. This feature is visible as a small peak in the early spectra, but it clearly develops a P-Cygni profile by day $28$. On day $8$ we also observe a faint absorption structure at $\sim 5500$ \AA, which \citet{bose13} suggest to be a possible high velocity component of He~I. At the epoch of the $V$-band maximum (day $\sim 9$), the H$_\alpha$, H$_\beta$, H$_\gamma$ and H$_\delta$ lines are clearly visible. Typical Type IIP SNe metal lines are visible in the bluest part of the spectra after the $V$-band maximum, namely the Fe~II, Ti~II, Sc~II, Ba~II, and Ca~II H\&K features. As the ejecta expand (from day $24$), the continuum becomes weaker and redder in the UV-blue part of the spectra, while other lines appear redwards of $5000$ \AA. In particular, a strong Ca~II P-Cygni feature stands out at $\sim 8570$ \AA\  on day $23$, which at later epochs (see day $77$) deblends into the three Ca~II IR triplet components at $8498$ \AA, $8542$ \AA, and $8662$ \AA.

\begin{figure}  
\includegraphics[width=16.3cm]{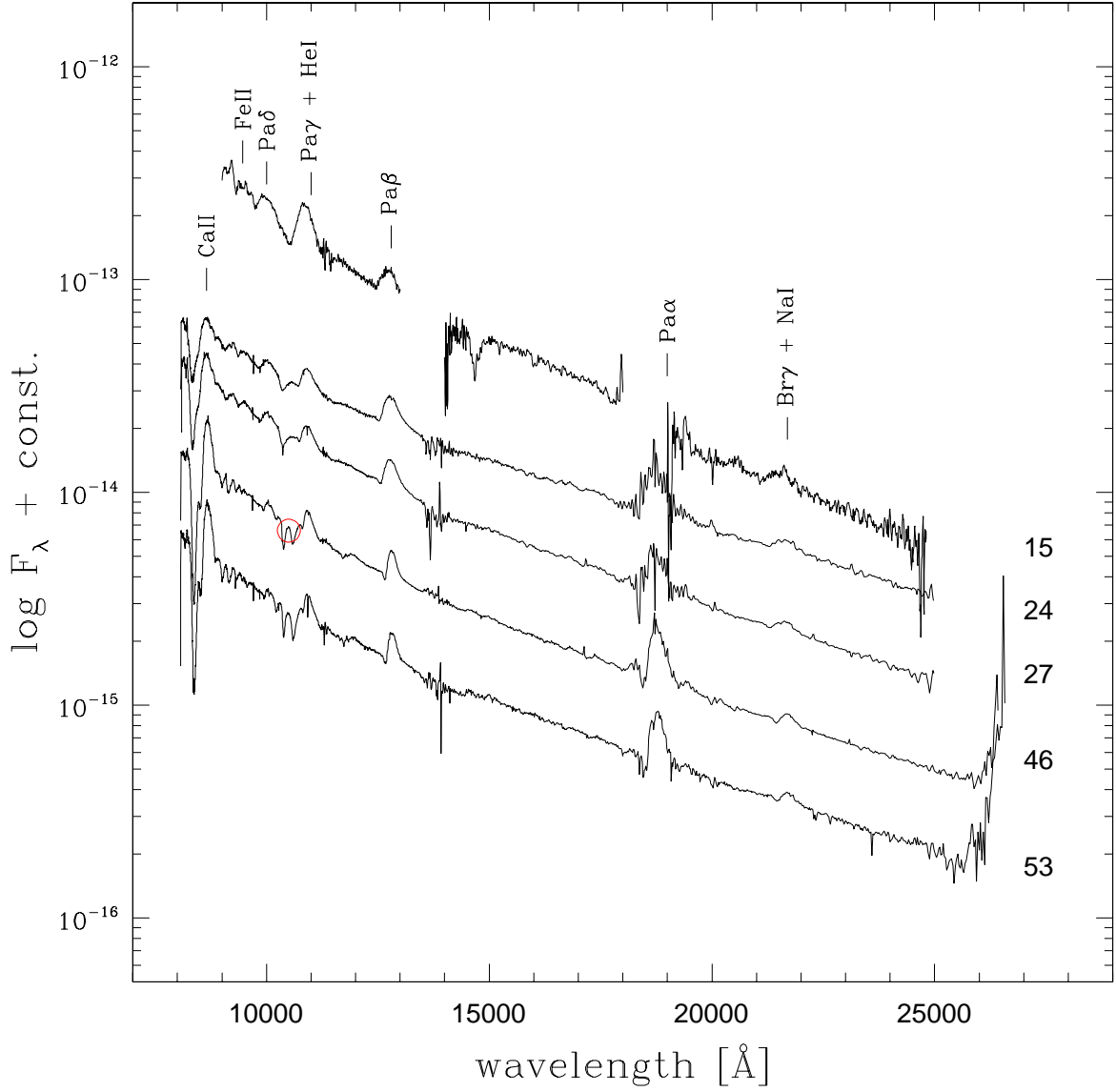} 
\caption{NIR spectra of SN 2012aw. Individual spectra have been shifted in flux, for clarity reasons. The most prominent features have been labelled. Numbers on the left indicate the epochs from the explosion. The red open circle marks the position of the unidentified feature discussed in the text.}
\label{nir_spectra} 
\end{figure}

Figure \ref{nir_spectra} shows the NIR spectroscopic evolution. The first
spectrum has been masked in the regions of low atmospheric transmission, since they appeared very noisy. Our time coverage ranges from day $15$ to day $53$. The H~I Paschen series is clearly visible at all reported phases, with Pa$_\gamma$ ($10938$ \AA) possibly blended with He~I ($10830$ \AA). A possible blend of the Brackett Br$_\gamma$ line with the Na~I is also visible in all spectra.  Redward of the Ca~II NIR triplet an Fe~II line is visible, which could be blended with Paschen Pa$_\epsilon$. Finally, we note the development of an unidentified P-Cygni line on day $24$ at $\sim 10400$ \AA. Searching for a possible identification we consulted the National Institute of Standards and Technology archive\footnote{http://www.nist.gov/pml/data/asd.cfm} and the \texttt{SYNOW} spectral synthesis code (e.g. \citealt{millard99}, \citealt{branch02}; \citealt{parrent07} for the \texttt{SYNOW} $2.0$ description), but could not find a reasonable match with usual ions showed by SNe. Therefore, we tentatively suggest that this is a high velocity Pa$_\gamma$ line. If this is the case, the Pa$_\gamma$ absorption clearly splits into two components ($10340$ \AA\  and $10560$ \AA) in the day $46$ spectrum, which would correspond to velocities of $\sim16000$ and $\sim 10000$ km s$^{-1}$, respectively. However, we do not see similar features for the other H lines.

\begin{figure}  
\includegraphics[width=16.3cm]{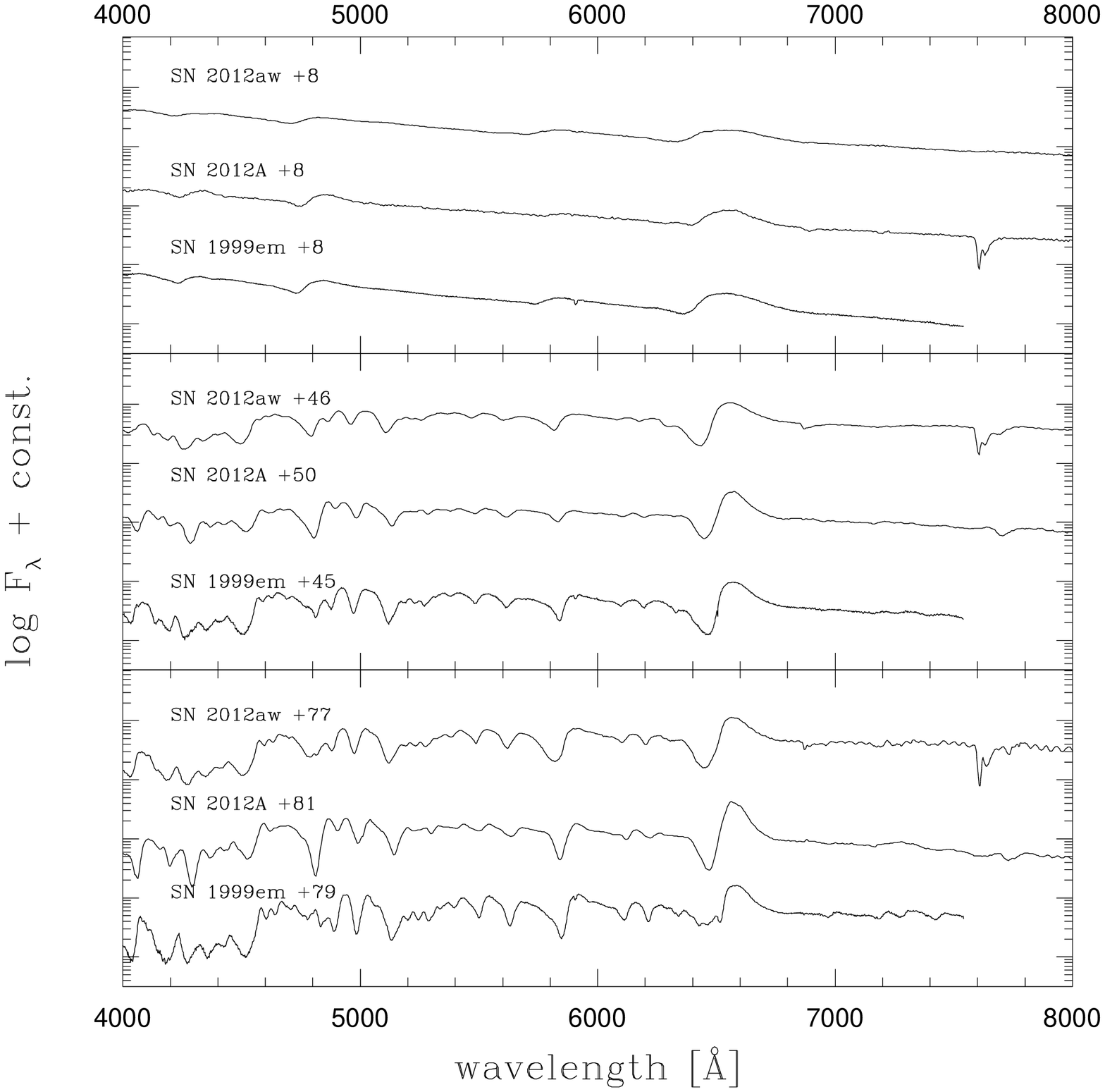} 
\caption{Comparison of the SN 2012aw spectra at selected phases with the Type IIP SNe SN199em and SN 2012A at similar phases. Top panel: about a week after explosion; middle panel: at middle plateau phase ($\sim 50$ days); bottom panel: in advanced plateau phase ($\sim 80$ days).}
\label{spec_comparison} 
\end{figure}

In Figure \ref{spec_comparison} we compare the spectra of SN 2012aw at various phases (around maximum, at about the middle of the plateau phase, and at the advanced plateau phase), with those of other well-studied Type IIP SNe, namely SN 1999em \citep{elmhamdi03} and SN 2012A \citep{tomasella13}. All the spectra were corrected to the rest wavelength and for the reddening. The spectra at all phases are very similar, with a blue continuum at early phases, the subsequent development of the typical Balmer lines, and the emergence of metal lines, at about one month.

\section{Physical Parameters}\label{phys}

\subsection{Bolometric light curve and $^{56}$Ni mass} \label{bol}

\begin{figure}  
\includegraphics[width=16.3cm]{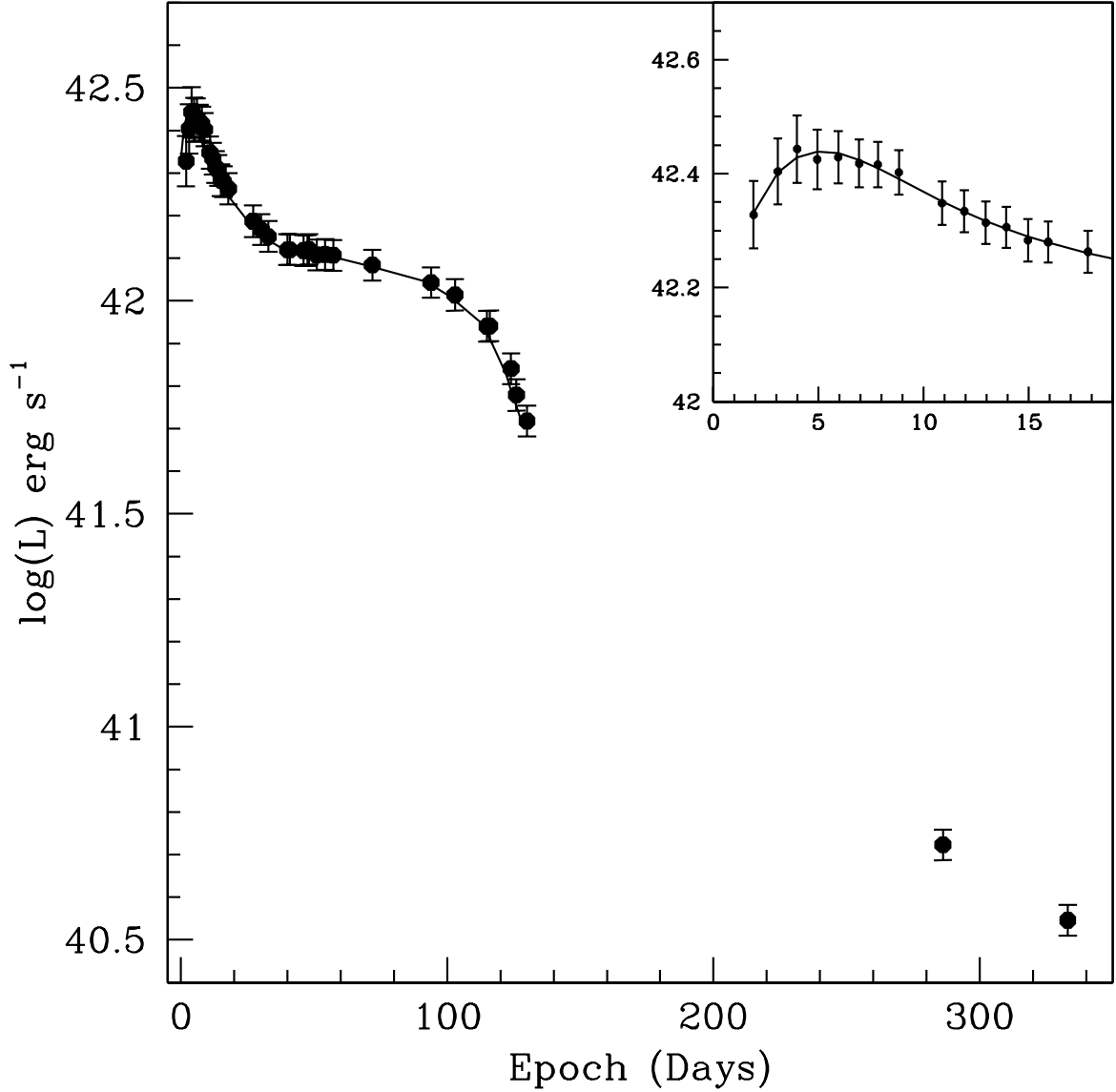} 
\caption{$uvoir$ bolometric light curve of SN 2012aw. The bolometric luminosity was obtained from a full set of \textit{Swift} $uvw2$, $uvw1$, Johnson-Cousins $UBVRI$ and near-infrared $JHK$ measurements, following the procedure described in the text. Errorbars are generally negligible with respect to the size of the plotted points. The inset shows a zoom of the first $20$ days.} 
\label{bolom} 
\end{figure}

A bolometric light curve (Figure \ref{bolom}) was obtained by integrating our
photometric measurements and the \textit{Swift} UV photometry \citep{bayless13}, and using the above adopted reddening and the distance
modulus. We converted $uvw2uvw1UBVRIJHK$ magnitudes\footnote{We did not use the \textit{Swift} uvm2 band, due to the lower number of measurements available and to the higher photometric errors in this band.} into monochromatic fluxes at the effective wavelength of the filter, then corrected these fluxes for the adopted extinction according to the extinction law from \citet{cardelli89}, and finally  integrated the resulting SED over the range of wavelength, after assuming zero flux at the integration limits. We estimated the flux only for the phases in which $V$-band observations were available. The photometric data in the other bands were estimated at these phases by interpolating magnitudes in adjacent nights. Finally, flux was converted into luminosity using the adopted distance modulus. The peak of the bolometric luminosity is reached at day $\sim 4$ at a luminosity of $L_{bol}=(2.8 \pm 0.5) \times 10^{42} $ erg s$^{-1}$. In Figure \ref{bolom} we also show a close-up of the evolution of the bolometric luminosity during the first $20$ days. The maximum is quite sharply reached, followed by a decline with a sort of flattening, and by a change in the decline slope at day $\sim 9$. The latter coincides with the already discussed feature in the $I$ band (see Sect. \ref{phot_evol}).

\begin{figure}  
\includegraphics[width=16.3cm]{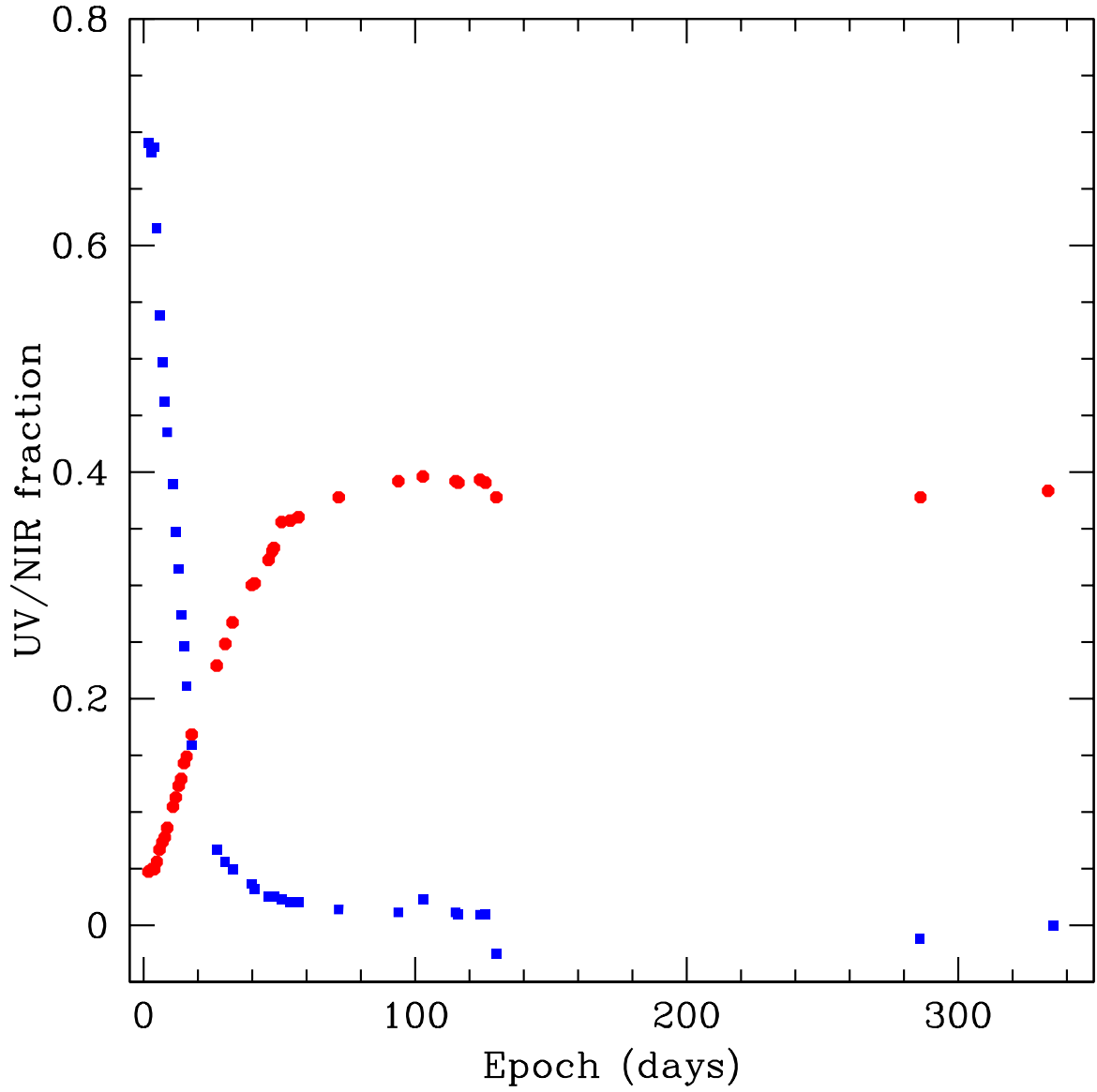} 
\caption{UV (blue filled squares) and NIR (red filled circles) contribution to the total flux.}
\label{flux_perc} 
\end{figure}

\begin{figure}  
\includegraphics[width=16.3cm]{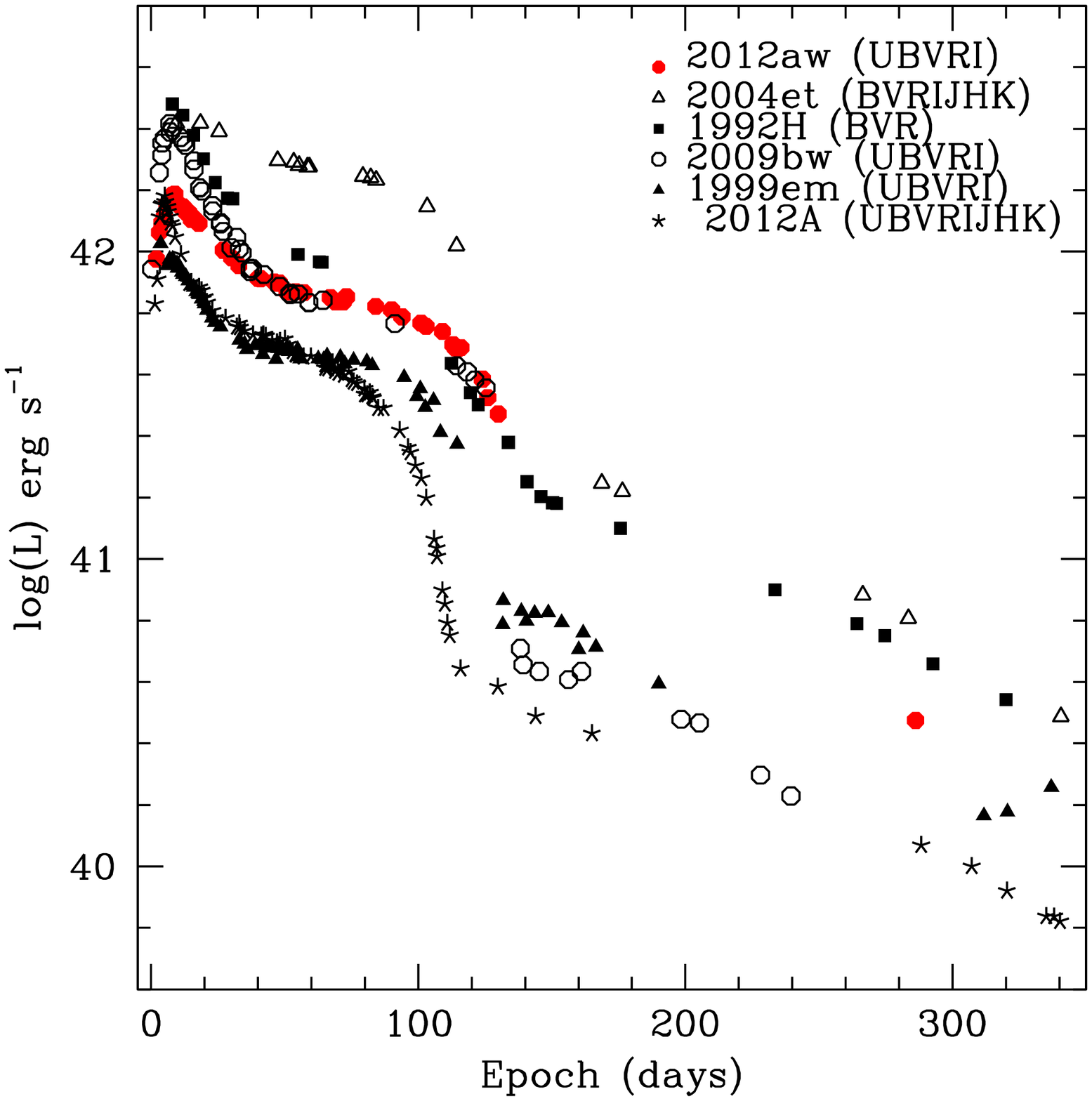}
\caption{$UBVRI$ Pseudo-bolometric light curve of SN 2012aw. The light curve is
compared with the Type IIP SNe SN 1992H, SN 1999em, SN 2004et, SN 2009bw, and SN 2012A.} 
\label{bolom_comparison}
\end{figure}

Taking advantage of our full UV-optical-NIR ($uvoir$) dataset, in Figure
\ref{flux_perc} we show the contribution of the \textit{Swift} $uvw2$, $uvw1$ bands (filled squares) and of the NIR bands (filled circles) to the total flux. The NIR contribution shows a progressive rise during the photospheric phase up to the end of the plateau, and then remains approximately constant
during the nebular phase, at least until day $\sim330$. This behaviour is similar to other Type IIP SNe, such as SN 2004et \citep{maguire10} and SN 2007od \citep{inserra11}. The UV contribution steeply decreases after the explosion, showing a ``knee'' at the beginning of the plateau. By the middle of the plateau it decreases to the $2\%$ level of the total flux at the middle of the plateau, and becomes negligible ($\lesssim 1\%$) at the end of the photospheric phase. In order to compare SN 2012aw with other SNe found in the literature, for which only a limited wavelength coverage was available, we also calculated a $UBVRI$ pseudo-bolometric light curve of SN 2012aw. The comparison of SN 2012aw with SN 1992H \citep{clocchiatti96}, SN 1999em \citep{elmhamdi03}, SN 2009bw \citep{inserra12}, SN 2004et \citep{maguire10} and SN 2012A \citep{tomasella13} in Figure \ref{bolom_comparison} shows that SN 2012aw belongs to the bright branch of the luminosity distribution of Type IIP events. The $^{56}$Ni mass was estimated by comparing the luminosity of SN 2012aw with that of SN 1987A during the nebular phase, assuming a similar $\gamma-$ray deposition fraction such that: 

\begin{equation} 
M(^{\rm{56}Ni}_{\rm{12aw}})=M(^{\rm{56}Ni}_{\rm{87A}}) \times \frac{L_{\rm{12aw}}}{\rm{L_{87A}}} M_\odot \end{equation}

where the luminosities must be compared at similar epochs. We adopted for SN
1987A a $^{56}$Ni mass of $M(^{\rm{56}Ni}_{\rm{87A}})=0.073 \pm 0.012$ $M_\odot$, which is the weighted mean of the values given by \citet{arnett89b} and by \citet{bouchet91}, and the ultraviolet-optical-infrared bolometric luminosity given by \citet{bouchet91}. We therefore obtained $M(^{\rm{56}Ni}_{12aw})=0.056 \pm 0.013$  $M_\odot$, as an average of the individual estimates at days $286$ and $333$. This value is in agreement, within the uncertainties, with the estimate of $0.06 \pm 0.01$ $M_\odot$ given by \citet{bose13}, obtained with the same method, and with the $0.062$ $M_\odot$, estimate of \citet{jerkstrand13}, based on the spectral synthesis models of the nebular phase.

The estimated nickel mass can be compared with the values inferred for our  SN sample, which range from $\sim 0.02  M_\odot$ (SN 1999em, \citealt{elmhamdi03}; SN 2009bw, \citealt{inserra12}) to $\sim 0.06 M_\odot$ (SN 2004et, \citealt{maguire10}) and $\sim 0.07 M_\odot$ (SN 1992H, \citealt{clocchiatti96}). These estimates, adopted from the original papers,
were derived using the same method as we follow for SN 2012aw, except for SN 1992H, whose $^{56}$Ni mass was estimated from a theoretical light curve.

\subsection{Expansion velocity, black body temperature and SED evolution}
\label{bb}

\begin{figure}  
\includegraphics[width=16.3cm]{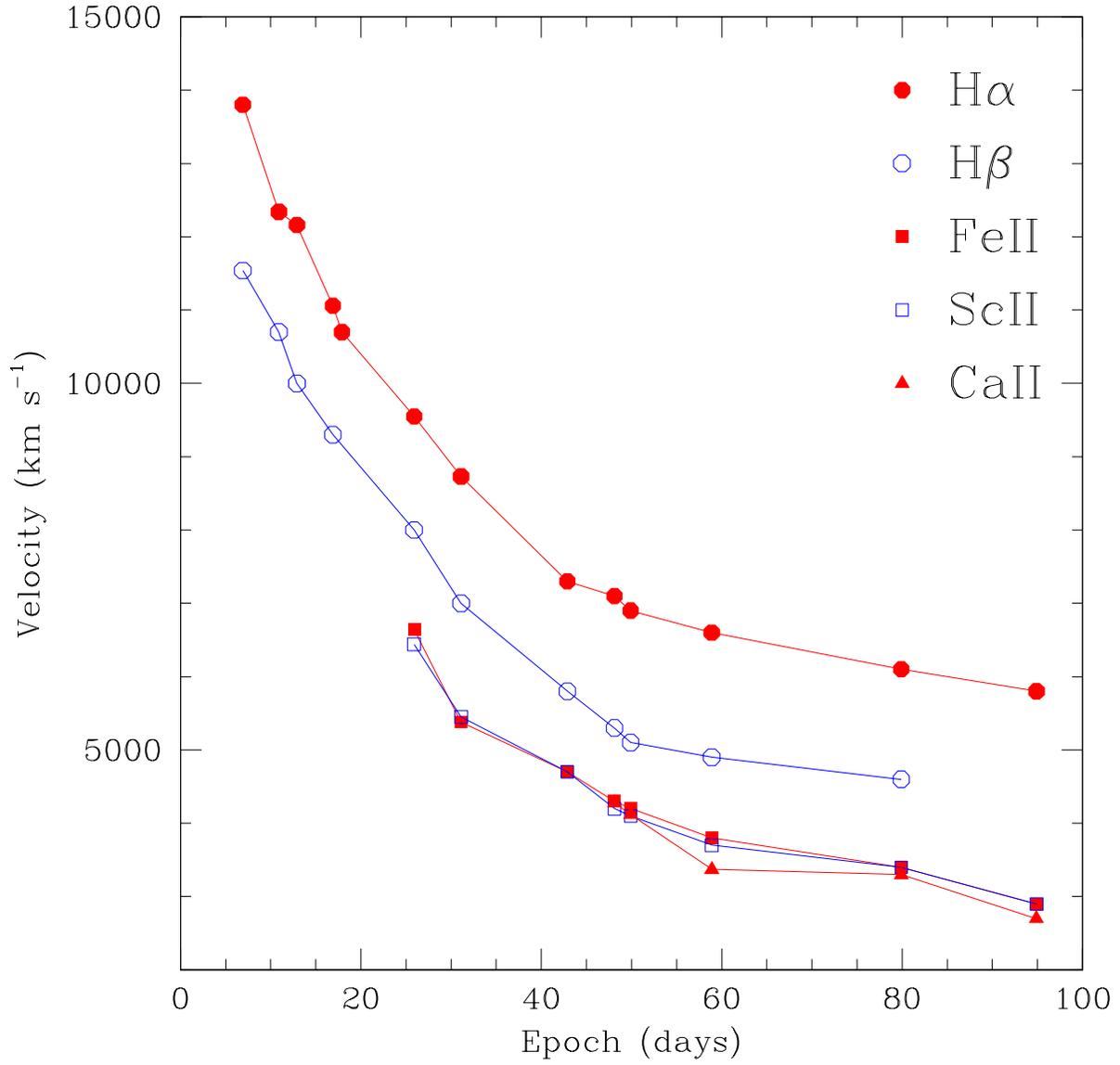} 
\caption{Line velocity evolution, estimated from the Doppler shift of the absorption minima, of H$\alpha$, H$\beta$, FeII (5169), ScII (6256), and CaII (8520).}
\label{vel_all} 
\end{figure}

Figure \ref{vel_all} shows the evolution of the photospheric expansion
velocities measured from the Doppler-shift of absorption minima of the H$\alpha$, H$\beta$ Fe~II ($5169$ \AA), Sc~II ($6245$ \AA) and Ca~II ($8520$ \AA) lines. Measurements have been performed by fitting the lines with a single gaussian profile. The H$\alpha$ and H$\beta$ lines are characterized by the highest velocities, starting from $\sim 14000$ and $\sim 12000$ km s$^{-1}$ on day $15$, respectively. Their velocities rapidly decrease and, at about $50$ days from the explosion, they reach an almost constant value of $\sim 7000$ and $\sim 5000$ km s$^{-1}$, respectively. We note that these values appear larger than in other Type IIP SNe at similar phases, (e.g. SN 2012A, \citealt{tomasella13}, their Figures 12 and 13; SN 2009bw, \citealt{inserra12}, their Table 9; SN 2004et, \citealt{maguire10}, their Figure 20). As is typical in Type IIP SNe, H$\alpha$ and H$\beta$ velocities are higher, since these spectral features are formed at larger radii than those of most metal lines. The Fe~II and Sc~II velocities are considered to be better tracers of the photospheric velocity, since the relevant transitions have small optical depths. They show a behaviour very similar to each other, both settling to $\sim 3000$ km s$^{-1}$ after about two months. Other luminous Type IIP SNe such as SN 2009bw \citep{inserra12}, SN 2004et \citep{maguire10}, and SN 1999em \citep{elmhamdi03} exhibit similar line velocities, while those of SN 2005cs appear lower (see \citealt{maguire10}, their Figure 21). The velocity evolution of the Ca~II feature resembles that of the Fe~II and Sc~II lines, but with a slightly larger scatter, due to measurement uncertainties.

\begin{figure}  
\includegraphics[width=16.3cm]{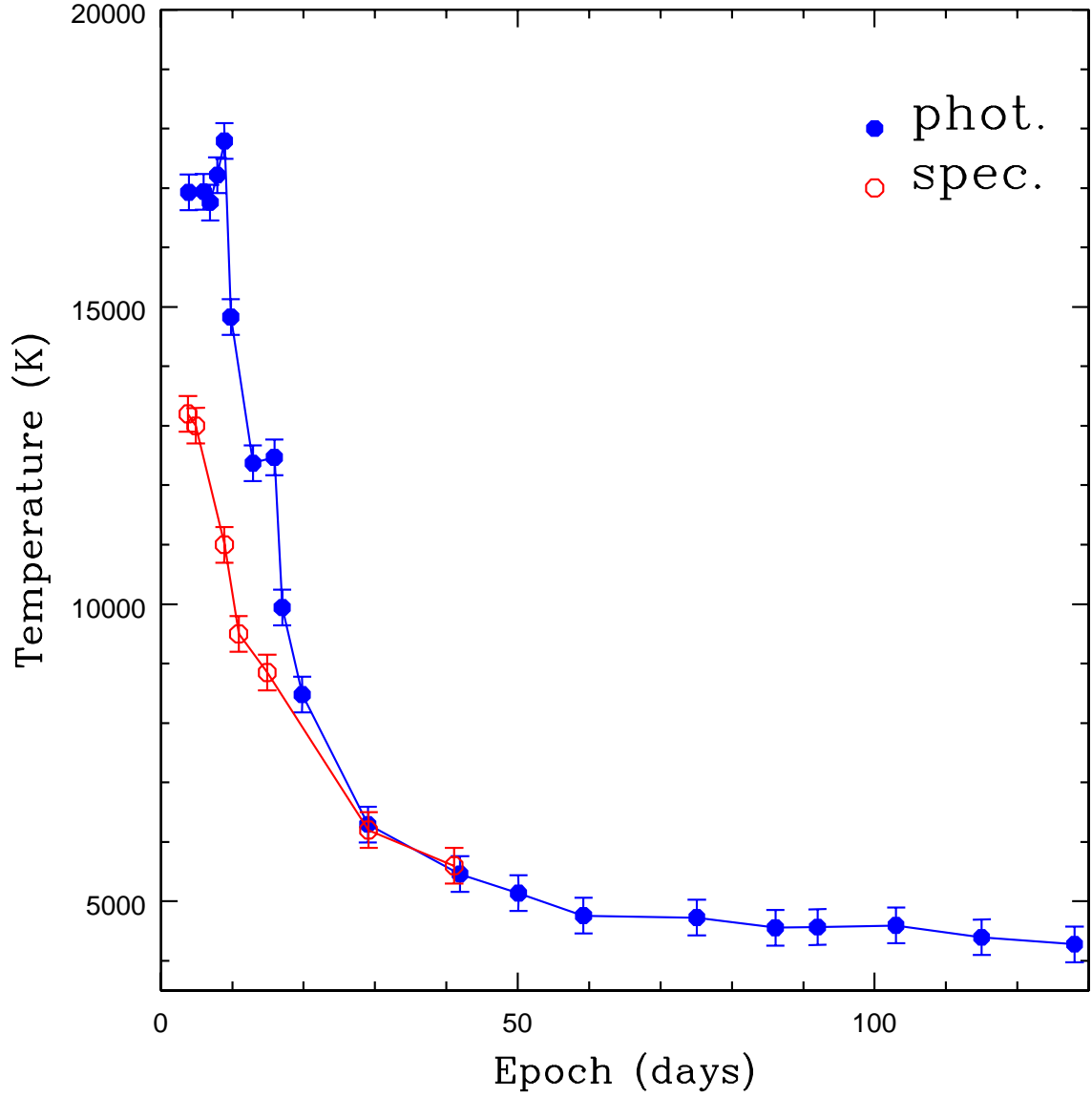}
\caption{Temperature evolution of SN 2012aw, derived from blackbody fits to the
observed fluxes in the range from the \textit{Swift} $uvw2$- to the $K$-bands
(blue filled circles) and from the continuum of selected spectra (red open
circles).} 
\label{temp_all} 
\end{figure}

Figure \ref{temp_all} shows the time evolution of the photospheric temperature, evaluated with a blackbody fit to the photometric data (blue filled circles) and to the spectral continuum (red open circles). In the first $\sim 20$ days, photometry-based temperatures appear systematically hotter than the spectral-based measurements, while on day $25$, the measurements agree within the uncertainties. A possible explanation of this behaviour is that our spectra do not include the ultra-violet wavelengths covered by the \textit{Swift} photometry. The evolution of the spectral continuum temperature looks similar to that in other Type IIP SNe (e.g. \citealt{inserra12}, their Figure 11). Interestingly, between day $\sim 12$ and day $\sim 16$ a small plateau in the temperature evolution is visible. The same feature is also visible in \citet{bose13}, their Figure 7, and it is also suggested in our individual light curves, already discussed. This is in correspondence with the light curve plateau transition (see Figure \ref{vri_zoom}). Finally, we note that Figure \ref{temp_all} shows an almost constant temperature from day $\sim 30$, in agreement with the \citet{bayless13} findings for SN 2012aw.

\begin{figure} 
\includegraphics[width=16.3cm]{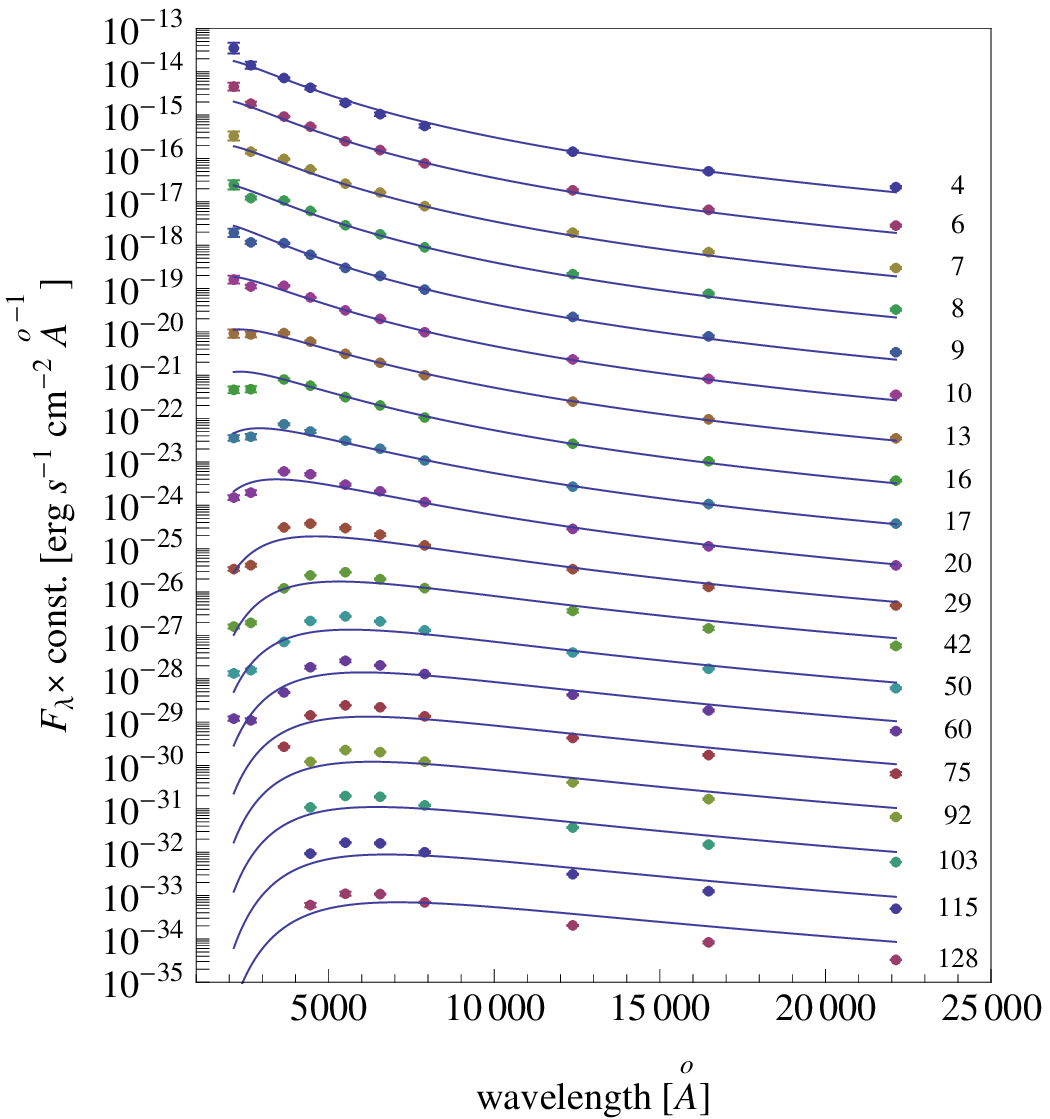} 
\caption{Time evolution of the spectral energy distribution of SN 2012aw. Filled circles depict the fluxes at the effective wavelengths of the photometric filters. Solid lines show the blackbody continuum fits. Numbers indicate the phases from core-collapse. Observed fluxes have been corrected for the adopted extinction and distance.} 
\label{sed_evol} 
\end{figure}

Figure \ref{sed_evol} shows the SED evolution between day $\sim 4$ and day
$\sim 132$. Our SED was based on our optical-NIR photometry, complemented with \textit{Swift} UV $uvw2$ and $uvw1$ data \citep{bayless13} which cover approximately the first $60$ days after the explosion. The wavelength
coverage ranges between $\sim 2000$ \AA\  to $\sim 22000$ \AA. Superimposed to the points are, for each epoch, blackbody continuum fits. During this time, the optical-NIR fluxes in the range $\sim 4000 - 22000$ \AA\ well resembles single blackbody curves. 

\section{Explosion and progenitor parameters}\label{model}  

Some observational quantities, namely the bolometric luminosity, the length of the plateau, and the evolution of line velocities and continuum temperature at the photosphere can be used to constrain the relevant physical parameters of the SN, that is the ejected mass, the progenitor radius, the explosion energy and the amount of $^{56}$Ni (e.g. \citealt{litvinova85}; \citealt{zampieri03}; \citealt{kasen09}).

We estimate these physical parameters for SN 2012aw by performing a
simultaneous $\chi^{2}$ fit of the aforementioned observational quantities
against model calculations, using the same well-tested procedure adopted for
modelling other core-collapse SNe (CC-SNe; e.g.~SNe 2007od, 2009bw, 2009E, and 2012A; see \citealt{inserra11}, \citealt{inserra12}, \citealt{pastorello12}, and \citealt{tomasella13}).

Two codes have been used to calculate the models: the  semi-analytic code
described in \citet{zampieri03} and the radiation-hydrodynamics code described in \citet{pumo10} and \citet{pumo11}. The first one solves the energy balance equation for a spherically  symmetric, homologously expanding envelope with constant density. It is used to perform preparatory  studies aimed at narrowing down the parameter space describing the CC-SN progenitor at the explosion and, consequently, to guide the more realistic but time consuming simulations performed  with the radiation-hydrodynamics code. This code is able to simulate the evolution of the physical properties of the  CC-SN ejecta and the evolution of the main CC-SN observables up to the nebular stage, solving the  equations of relativistic radiation hydrodynamics for a self-gravitating fluid which interacts with  radiation. The main features of this code are: i) a fully implicit Lagrangian approach to the solution of the system of relativistic radiation  hydrodynamics equations, ii) an accurate treatment of radiative transfer coupled with relativistic  hydrodynamics, and iii) a self-consistent treatment of the evolution of ejected material taking into  account both the gravitational effects of the compact remnant and the heating effects due to decays  of radioactive isotopes synthesized during the CC-SN explosion.

We point out that our modelling using both the aforementioned codes is
appropriate only if the emission from the CC-SN is dominated by the thermal
balance in the expanding ejecta. In the case of SN 2012aw, there could be
contamination from an early interaction with circumstellar matter (see Sect.
\ref{intro}), which may partially affect the  observables during the early
post-explosion evolution (first $\sim 30$ days after explosion).  Nevertheless, since there is no evidence that such contamination continues and dominates during most of the evolution, we assume that our modelling can be applied to SN 2012aw and returns a robust estimate of  the physical properties of the progenitor (as already done for other CC-SNe  with possible contamination from a relatively ``weak'' interaction like SNe 2007od and 2009bw; see \citealt{inserra11,inserra12}). However, in the $\chi^{2}$ fit we do not include the data taken at early  phases because the behaviour of the observational quantities could be contaminated by a possible interaction. In addition, during such phases there is significant emission from the outermost shell of the ejecta, which is accelerated to very high velocities and is not in homologous expansion \citep{pumo11}. The structure, evolution and emission properties of this shell are not well reproduced in our simulations because at present we adopt an \textit{ad hoc} initial density profile, not one consistently derived from a post-explosion calculation.

The explosion epoch and distance modulus adopted here are those reported in Sect. ~\ref{intro} and Sect. ~\ref{m95}, respectively. A $^{56}$Ni mass of $\sim 0.06$ $M_\odot$ is assumed (see Sect. ~\ref{bol}). 

\begin{figure} \includegraphics[width=16.3cm]{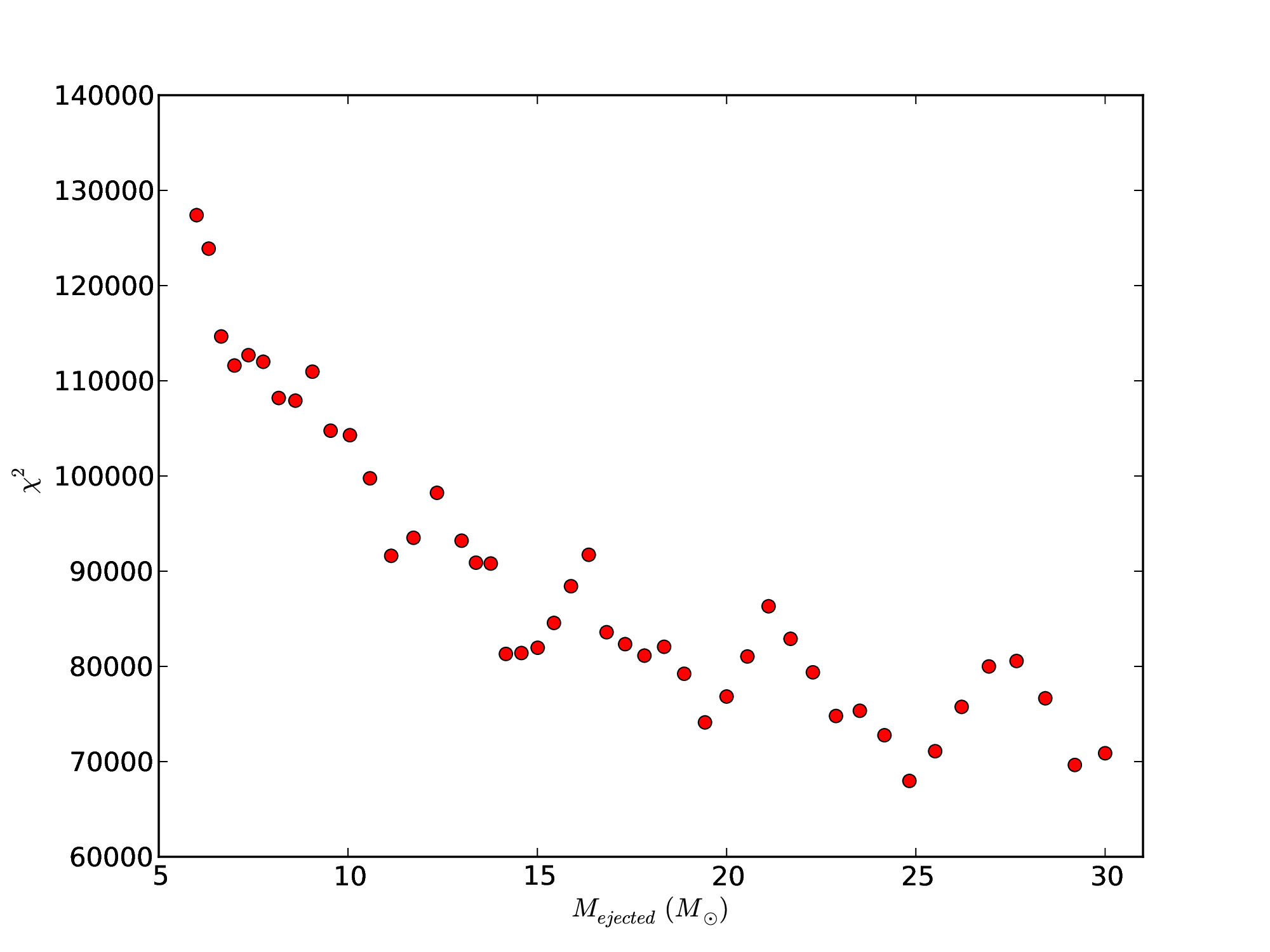}
%\includegraphics[width=8.3cm]{output_1767.ps} 
%\includegraphics[width=8.3cm]{output_1400.ps} 
%\includegraphics[width=8.3cm]{output_1065.ps}  
%\caption{Semi-analytical modelling for the models $1767$, $1400$ (left) and $1065$ (right), corresponding to an ejected mass of $\sim 12 M_\odot$, $\sim 16 M_\odot$ and $\sim 21M_\odot$, respectively. Estimated physical parameters are shown in the panels.}
\caption{$\chi^2$ distribution of the fit of the semi-analytical model to the observed quantities, as a function of the estimated ejected mass.}
\label{models_comparison} \end{figure}

\begin{figure}  
\includegraphics[angle=-90,width=16.3cm]{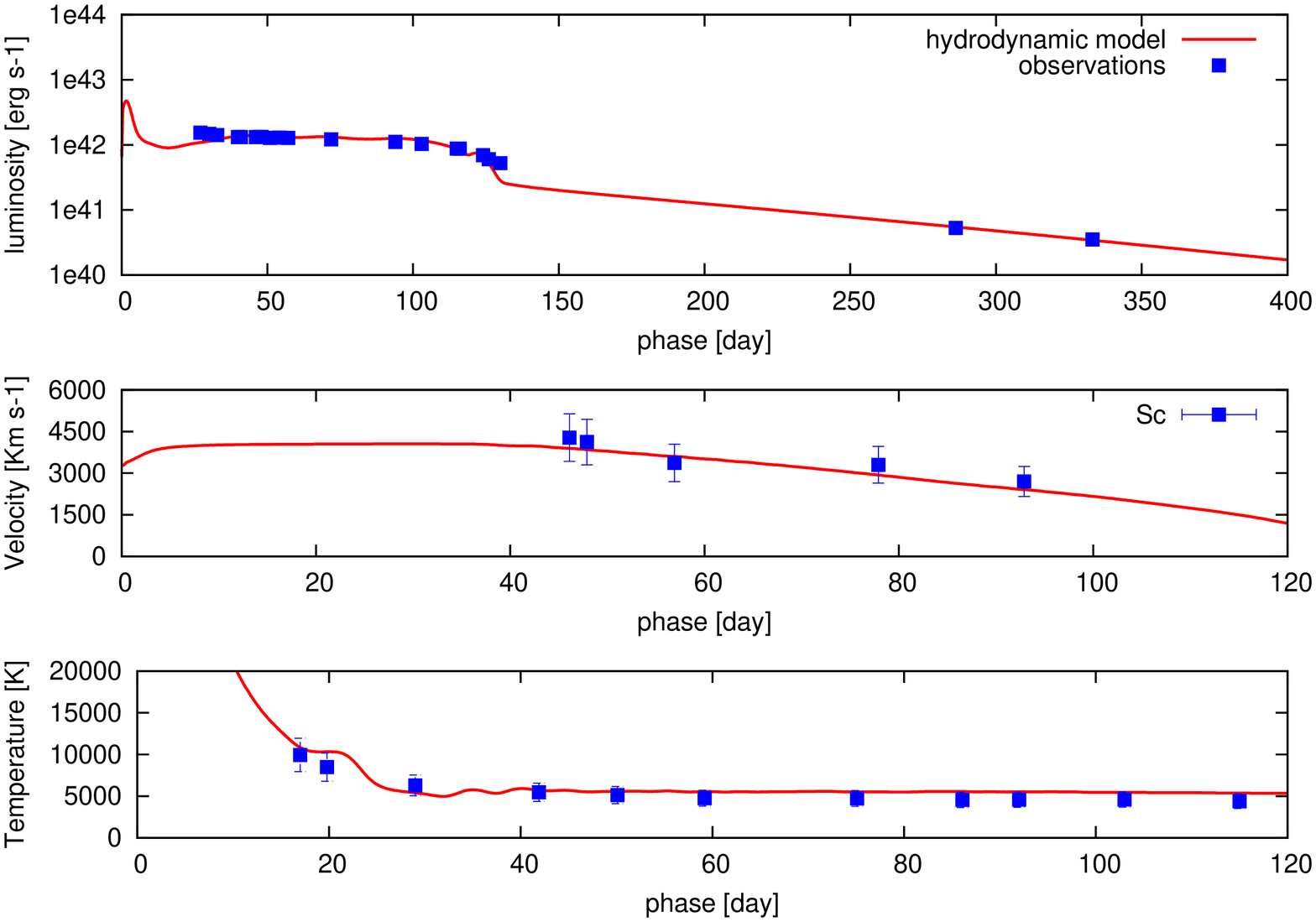}
%\caption{Comparison of the evolution of the main observables of SN 2012aw with the best-fit models computed with the semi-analytic code (total energy $\sim 1.6$ foe, initial radius $\sim 2.2 \times 10^{13}$ cm, envelope mass $\sim 21 M_\odot$) and with our radiation-hydrodynamics code (total  energy $1.7$ foe, initial radius $4 \times 10^{13}$ cm, envelope mass $23 M_\odot$). Top, middle, and bottom panels show the bolometric light curve, the photospheric velocity, and the photospheric temperature as a function of time. To better estimate the photosphere velocity from observations, we use the minima of the profile of the Sc~II lines.} \label{modelling}  \end{figure}
\caption{Comparions of the evolution of the main observables of SN 2012aw with the best-fit model computed with our radiation-hydrodynamics code (total  energy $1.5$ foe, initial radius $3 \times 10^{13}$ cm, envelope mass $19.6$ $M_\odot$). Top, middle, and bottom panels show the bolometric light curve, the photospheric velocity, and the photospheric temperature as a function of time. To better estimate the photosphere velocity from observations, we use the minima of the profile of the Sc~II lines.} \label{modelling}  \end{figure}

We computed an extended grid of semi-analytical models, covering a significant range in mass. In Figure \ref{models_comparison} we show the $\chi^2$ of the models as a function of the ejected mass. The distribution has a broad structured minimum extending from $\sim 15$ to $\sim 28$ $M_\odot$. Significant local minima occur at $\sim 16$ $M_\odot$, $\sim 19$ $M_\odot$, and $\sim 25$ $M_\odot$, while an additional less prominent minimum occurs at $\sim 12$ $M_\odot$. We explored the minima at $\sim 19$ and $\sim 25$ $M_\odot$ to constrain the parameter space for the radiation-hydrodynamics simulations. The latter were run varying the ejected mass in the range $16-27$ $M_\odot$ and are in fair agreement with the semi-analytical models. Figure \ref{modelling} shows the result for the best fitting semi-analytical and hydrodynamical simulations, giving an ejected mass of $\sim 20$ $M_\odot$, a total (kinetic plus thermal) energy of $1.5$ foe and an initial radius of $3 \times 10^{13}$ cm. These values are consistent with a scenario where the SN is produced by a relatively standard explosion of a supergiant progenitor with a total mass of $\sim 21$ $M_\odot$ at explosion. We note that the local minimum of the $\chi^2$ at $\sim 16$ $M_\odot$ is close to the $\sim 15$ $M_\odot$ estimate of the progenitor mass given by \citet{kochanek12} and \citet{bose13}, and to the $17-18$ $M_\odot$ value given by \citet{vandyk12}. However, with an ejected mass of $\sim 15$ $M_\odot$ our radiation-hydrodynamics code fails to reproduce all the observed features. In particular, it is not possible to reproduce at the same time the observed expansion velocity and the length of the plateau, which are diagnostics that are basically independent of the adopted reddening and distance. As a matter of fact, when adopting the high reddening estimate $E(B-V)=0.19$ mag discussed above, the same procedure gives an ejected mass of $\sim 21 - 23$  $M_\odot$, a total energy of $1.6 - 1.7$ foe, and initial radius of $2-4 \times 10^{13}$ cm.

\section[]{Discussion and Conclusions}\label{conclusions}  

We have presented the results of our photometric and spectroscopic campaign of the Type IIP SN 2012aw. Our photometry maps the SN from the explosion up to the end of the plateau (at day $\sim 125$), in the UV-optical-NIR bands. Moreover, two additional epochs were collected in the nebular phase (at day $286$ and day $333$), to get an estimate of the $^{56}$Ni mass. Spectroscopic data map the SN evolution from day $2$ to day $90$. Our data allowed us to draw a detailed picture of SN 2012aw, by deriving all the relevant diagnostics, namely the expansion velocity and photospheric temperature evolution, and estimating its physical parameters. We adopt the distance modulus ($\mu=29.96 \pm 0.04$ mag) by averaging the Cepheids \citep{freedman01} and the TRGB \citep{rizzi07} distances, while estimating the Galactic reddening from \citet{schlegel98}. The host reddening was evaluated by measuring the EW(Na~ID) on a high-resolution spectrum, and adpting the \citet{poznanski12} calibration we derived $E(B-V)=0.058 \pm 0.016$ mag. Taking into account a foreground reddening of $E(B-V)= 0.028$ mag, estimated from the \citet{schlegel98} maps, we end up with the total reddening (foreground and host) $E(B-V)=0.086 \pm 0.02$ mag.

With the adopted distance and reddening values, our analysis of the bolometric light curve shows that SN 2012aw belongs to the high branch of Type IIP SNe luminosities and allows us to estimate an ejected $^{56}$Ni mass of $ \sim 0.056 \pm 0.013$ $M_\odot$. The SED shows a generally good fit with a single blackbody curve. 

From the collected spectra we measure a fairly large initial expansion velocity, of $\sim 14,000$ km s$^{-1}$ in the H$\alpha$ line. After $\sim 50$ days from the explosion, the H$\alpha$ and H$\beta$ lines settle on a constant value of $\sim 6000$ and $\sim 5000$ km s$^{-1}$, respectively. Starting from day $\sim 25$, we obtain an expansion velocity of $\sim 3000$ km s$^{-1}$ from the Fe II and Sc II lines, which are known to be better tracers of the photospheric velocities. This behaviour is in agreement with those shown by other luminous Type IIP SN, such as SN 2009bw \citep{inserra12}.

We estimate the physical parameters of SN 2012aw and its progenitor by means of the hydrodynamical modelling described in Sect. \ref{model}, which uses the radiation-hydrodynamics code (\citealt{pumo10}; \citealt{pumo11}). Our
simulations suggest that the envelope mass is $M_{env} \sim 20$ $M_\odot$, the radius is $R \sim 3 \times 10^{13}$ cm, the energy is $E \sim 1.5$ foe, and the initial $^{56}$Ni in the $\sim 0.05 - 0.06$ $M_\odot$ range. We explicitly note that our progenitor mass and radius estimates are in fair agreement with the independent evolutionary model-based values of \citet{fraser12} based on a direct progenitor detection: $M_{ZAMS} \sim 14 - 26$ $M_\odot$ and $R > 500$ $R_\odot \simeq 3.5\times 10^{13}$ cm. Taken at face value, these estimates indicate a massive SN progenitor, with a mass significantly higher than the observational limit of $16.5 \pm 1.5$ $M_\odot$ that raised the ``RSG problem'' \citep{smartt_etal_09}, thus is in good agreement with the higher mass limit of $21^{+2}_{-1}$  $M_\odot$ found by \citet{walmswell12}. However, our values are considerably larger than those estimated by \citet{kochanek12}, $L < 10^4 L_\odot$, $M < 15 M_\odot$, obtained by carefully modelling the circumstellar extinction and not simply assuming an interstellar extinction law for the circumstellar dust. Moreover, it has been reported in the literature that the ejecta masses estimated from the modelling are generally too high to be consistent with the initial masses determined from direct observations of SN progenitors (e.g. \citealt{utrobin09}, \citealt{maguire10}). However, the code used here gives lower ejecta masses, as also noted in \citet{jerkstrand12}. 
It is interesting to compare our results with those obtained by \citet{bose13}, who give an estimate of the explosion energy and the progenitor mass by using the analytical relations given by \citet{litvinova85} and adopting the radiation hydrodynamical simulations provided by \citet{dessart10}. Their analysis points toward an explosion energy in the range $1-2$ foe and a progenitor mass in the $14-15$ $M_\odot$ range. It should be noted that \citet{bose13} found several similarities between SN 2012aw and SN 2004et and SN 1999em, on the basis of \citet{utrobin09} and \citet{utrobin11} investigations. However, in the same papers the estimated progenitor masses are quite large, of the order of $20-25$ $M_\odot$. Moreover, \citet{bose13} found some evidence of interaction with the circumstellar medium, which could imply a large mass loss during the progenitor star's lifetime too large to be reconciled with a star of initial mass of $14-15 M_\odot$. Clearly, such differences are due mostly to the different models adopted and to the fact that there is still an issue regarding reconciling progenitor masses (which are model dependent) with ejecta masses (also model dependent). Therefore, it would be interesting to perform a detailed comparison of the different available codes on the same objects, to check how consistent the results are.\footnote{A similar experiment was already performed to test how different evolutionary models could determine the star formation histories of resolved stellar populations (the \textit{Coimbra Experiment}, see \citealt{skillman02}).} It should also be noted that the analysis of the nucleosynthesis products of SN 2012aw performed by \citet{jerkstrand13} seems to rule out a high-mass progenitor, in that the observed lines consistent with a progenitor in the $14 - 18$ $M_\odot$ range. However, as pointed out by the same authors, the link between progenitor mass and nucleosynthesis depends on some as yet uncertain processes in the input physics of the stellar evolution models, such as semi-convection, overshooting and rotation. Quoting \citet{jerkstrand13}: \textit{``Understanding the differences in results between progenitor imaging, hydrodynamical modeling, and nebular phase spectral analysis is a high priority in the Type IIP research field''}. Moreover, it is worth noting that, on the basis of our simulations, possible uncertainties in the local reddening do not have a dramatic impact on the estimate of the physical parameters of SN 2012aw. Indeed, when adopting the high reddening $E(B-V) = 0.19$ mag, our simulations give only slightly different values of the ejected mass, initial radius and explosion energy.

Finally, it should be noted that, as stated by \citet{brown13}: \textit{``the best we can say at the present time is what supernova mass limits might be consistent with observations. The idea of a limiting mass is itself an approximation, since the compactness of the core is not a monotonic function of main sequence mass [...], especially in the interesting range $20 - 35 M_\odot$''}.

\acknowledgments

We warmly thank our Referee for her/his helpful comments, that significantly improved the content and the readability of our manuscript.

This paper is based on observations collected at: the Copernico 1.82m Telescope and Schmidt 67/92 Telescope operated by the INAF - Osservatorio Astronomico di Padova at Asiago, Italy; the Galileo 1.22m Telescope operated by the Department of Physics and Astronomy of the University of Padova at Asiago; the 3.58m Italian Telescopio Nazionale Galileo (TNG) operated on the island of La Palma by the Fundaci\'{o}n Galileo Galilei of INAF (Istituto Nazionale di Astrofisica); the CNTAC, proposal CN2012A-103.

We acknowledge the TriGrid VL project and the INAF-Astronomical Observatory of Padua for the use of computer facilities. M.L.P.~acknowledges the financial support form the PRIN-INAF 2009 ``Supernovae Variety and Nucleosynthesis Yields'' (P.I.: S. Benetti) and PRIN-INAF 2011 ``Transient Universe: from ESO Large to PESSTO'' (P.I.: S. Benetti).

M.T.B., L.T., M.D.V., E.C., S.B., A.H., A.P., are partially supported by the
PRIN-INAF 2011 with the project ``Transient Universe: from ESO Large to PESSTO''.

G.P. and F.B. acknowledge support from ``Millennium Center for Supernova Science'' (P10-064-F), with input from ``Fondo de Innovación para la Competitividad del Ministerio de Economia, Fomento y Turismo de
Chile''.

G.P. acknowledges partial support by ``Proyecto interno UNAB DI-303-13/R''.

N.ER. and A.M.G acknowledge financial support by the MICINN grant AYA2011-24704/ESP, and by the ESF EUROCORES Program EuroGENESIS (MINECO grants EUI2009-04170).

S.T. acknowledges support by the Transregional Collaborative Research Centre TRR 33 ``The Dark Universe'' of the German Research Foundation.

The research leading to these results has received funding from the European research Council under the European Union's Seventh Framework Programme (FP7/2007-2013)/ERC Grant agreement n$^{\rm o} [291222]$ (P.I.: S.~J. Smartt).

\begin{table} \begin{minipage}{130mm}  \caption{Log of the spectroscopic
observations. For each spectrum, we list the UT observation date, the JD, the
epoch from the explosion, the wavelength range, the dispersion and the
instrument.} \label{log_spectra} %\vspace{0.5 cm}  \scriptsize
\begin{tabular}{lccccc}  \hline \hline Date        & JD         &  Epoch  & 
Range       &  Dispersion       &  Instrument    \\ dd/mm/yyyy  & $240000+$  & 
(days) &  \AA         &  \AA\  mm$^{-1}$  &                \\ 
\hline 
17/03/2012  & $56004.5$  &  $2.0 $ &  $3300-7800$  &  $169          $  &  Asiago1.2m + BC \\
19/03/2012  & $56006.6$  &  $4.1 $ &  $3300-7800$  &  $169          $  & Asiago1.2m + BC \\ 
19/03/2012  & $56006.7$  &  $4.2 $ &  $3200-9100$  & $220          $  &  NOT + ALFOSC    \\ 
20/03/2012  & $56007.6$  &  $5.1 $ &  $3300-7800$  &  $169          $  &  Asiago1.2m +BC  \\ 
20/03/2012  & $56007.8$  &  $5.3 $ &  $3000-8400$  &  $187          $  &  TNG + LRS         \\ 
20/03/2012  & $56007.8$  &  $5.3 $ &  $4500-10000$ &  $193          $  &  TNG + LRS       \\
21/03/2012  & $56008.6$  &  $6.1 $ &  $3300-7800$  & $169          $  &  Asiago1.2m + BC    \\ 
21/03/2012  & $56008.8$  &  $6.3 $ &  $4600-7900$  &  $61           $  &  TNG + SARG        \\ 
22/03/2012  & $56009.6$  &  $7.1 $ &  $3300-7800$  &  $169          $  &  Asiago1.2m + BC   \\ 
23/03/2012  & $56010.7$  &  $8.2 $ &  $3300-7800$  &  $169          $  &  Asiago1.2m + BC   \\
24/03/2012  & $56011.8$  &  $9.3 $ &  $3200-7000$  &  $185          $  &  CAHA2.2m + CAFOS     \\ 
25/03/2012  & $56012.7$  &  $10.2 $ & $3300-7800$  &  $169          $  &  Asiago1.2m + BC  \\ 
26/03/2012  & $56013.7$  &  $11.2 $ & $5000-11000$ &  $191          $  &  Ekar1.8m + AFOSC  \\ 
26/03/2012  & $56013.8$  &  $11.3 $ & $3500-7700$  &  $292          $  &  Ekar1.8m + AFOSC   \\ 
27/03/2012  & $56014.8$  &  $12.3 $ & $3500-7700$  &  $292          $  &  Ekar1.8m + AFOSC \\
28/03/2012  & $56015.8$  &  $13.3 $ & $3500-7700$  &  $292          $  & Ekar1.8m + AFOSC  \\ 
29/03/2012  & $56016.7$  &  $14.2 $ & $3300-7800$  &   $169          $  &  Asiago1.2m + BC     \\ 
29/03/2012  & $56016.8$  &  $14.3 $ & $5000-10100$ &  $95           $  &  TNG + SARG      \\ 
30/03/2012  & $56017.6$  &  $15.1 $ & $9000-14500$ &  $297          $  &  TNG + NICS      \\ 
30/03/2012  & $56017.6$  &  $15.1 $ & $14000-25000$&  $605          $  &  TNG + NICS      \\
30/03/2012  & $56017.8$  &  $15.3 $ & $3200-9100$  &  $220          $  & NOT + ALFOSC     \\ 
31/03/2012  & $56018.4$  &  $15.9 $ & $3300-7800$  &   $169          $  &  Asiago1.2m + BC   \\ 
31/03/2012  & $56018.6$  &  $16.1 $ & $3500-5200$  &  $64           $  &  WHT + ISIS        \\ 
31/03/2012  & $56018.6$  &  $16.1 $ & $5400-9500$  &  $120          $  &  WHT + ISIS        \\ 
02/04/2012  & $56020.3$  &  $17.8 $ & $3300-7800$  &  $169          $  &  Asiago1.2m + BC    \\
08/04/2012  & $56025.4$  &  $22.9 $ & $3000-8400$  &  $187          $  &  TNG + LRS         \\ 
08/04/2012  & $56025.4$  &  $22.9 $ & $4500-10000$ &  $193          $  &  TNG + LRS         \\ 
08/04/2012  & $56025.5$  &  $25.0 $ & $8000-25000$ &  $446          $  &  Magellan + FIRE  \\ 
11/04/2012  & $56028.6$  &  $26.1 $ & $8000-25000$ &  $446          $  &  Magellan + FIRE  \\ 
13/04/2012  & $56030.4$  &  $27.9 $ & $3700-9300 $ &  $185          $  &  NTT + EFOSC2        \\
25/04/2012  & $56043.5$  &  $41.0 $ & $3200-9100$  &  $220          $  &  NOT + ALFOSC     \\ 
30/04/2012  & $56047.6$  &  $45.1 $ & $8000-25000$ &   $446          $  &  Magellan + FIRE   \\ 
01/05/2012  & $56048.9$  &  $46.4 $ & $3700-9300 $ &  $185          $  &  NTT + EFOSC2      \\ 
07/05/2012  & $56054.5$  &  $52.0 $ & $8000-25000$ &  $446          $  &  Magellan + FIRE   \\ 
11/05/2012  & $56058.6$  &  $56.1 $ & $3200-9100$  &  $220          $  &  NOT + ALFOSC     \\
01/06/2012  & $56080.4$  &  $77.9 $ & $3200-9100$  &  $220          $  & NOT + ALFOSC     \\ 
16/06/2012  & $56095.4$  &  $92.9 $ & $3300-7800$  &  $169          $  &  Asiago1.2m + BC    \\

\hline \end{tabular} \end{minipage} \end{table}

\end{document}